\documentclass[prb,twocolumn,nofootinbib]{revtex4}
\usepackage{amssymb}
\usepackage{color}

\usepackage{amsthm}
\usepackage{graphicx}
\usepackage{dcolumn}
\usepackage{bm}
\usepackage{subfigure}
\usepackage{amssymb}
\usepackage{verbatim}
\usepackage{amscd}
\usepackage{amssymb}
\usepackage{amsmath, amsfonts}
\usepackage{setspace}
\usepackage[]{graphicx}        
\usepackage{amsthm}
\usepackage{enumerate}

\theoremstyle{plain}

\theoremstyle{plain}

\theoremstyle{plain}
 
\theoremstyle{plain}

\theoremstyle{remark}

\theoremstyle{conjecture}

\theoremstyle{observation}

\theoremstyle{definition}

\theoremstyle{corollary}

\theoremstyle{definition}

\theoremstyle{definition}

\theoremstyle{assumption}

\theoremstyle{definition}

\theoremstyle{problem}

\theoremstyle{fact}

\begin{document}

\title{Topological color code and symmetry-protected topological phases}
\author{Beni Yoshida}
\affiliation{Walter Burke Institute for Theoretical Physics and Institute for Quantum Information \& Matter, California Institute of Technology, Pasadena, California 91125, USA
}

\date{\today}

\begin{abstract}
We study $(d-1)$-dimensional excitations in the $d$-dimensional color code that are created by transversal application of the $R_{d}$ phase operators on connected subregions of qubits. 
We find that such excitations are superpositions of electric charges and can be characterized by fixed-point wavefunctions of $(d-1)$-dimensional bosonic SPT phases with $(\mathbb{Z}_{2})^{\otimes d}$ symmetry. 
While these SPT excitations are localized on $(d-1)$-dimensional boundaries, their creation requires operations acting on all qubits inside the boundaries, reflecting the non-triviality of emerging SPT wavefunctions. 
Moreover, these SPT-excitations can be physically realized as transparent gapped domain walls which exchange excitations in the color code. 
Namely, in the three-dimensional color code, the domain wall, associated with the transversal $R_{3}$ operator, exchanges a magnetic flux and a composite of a magnetic flux and loop-like SPT excitation, revealing rich possibilities of boundaries in higher-dimensional TQFTs. 
We also find that magnetic fluxes and loop-like SPT excitations exhibit non-trivial three-loop braiding statistics in three dimensions as a result of the fact that the $R_{3}$ phase operator belongs to the third-level of the Clifford hierarchy. 
We believe that the connection between SPT excitations, fault-tolerant logical gates and gapped domain walls, established in this paper, can be generalized to a large class of topological quantum codes and TQFTs.
\end{abstract}

\maketitle

\section{Introduction}

Classification of fault-tolerantly implementable logical gates in topological quantum error-correcting codes is an important stepping stone toward far-reaching goal of universal quantum computation~\cite{Gottesman98, Eastin09, Bravyi13b, Pastawski15}. Characterization of logical operators is also essential in understanding braiding and fusion rules of anyonic excitations arising in topologically ordered systems~\cite{Kitaev03, Levin05}. Although some classes of two-dimensional topological quantum codes are restricted to possess string-like logical operators only, there exist non-trivial topological quantum codes with two-dimensional logical operators. Namely, in the two-dimensional color code, transversal membrane-like phase operators lead to non-trivial action on the ground space which induce a non-trivial automorphism exchanging anyon labels~\cite{Bombin06, Bombin14}. Also, in three or more dimensions, the color code admits transversal non-Clifford logical phase gates, which are indispensable ingredient of fault-tolerant quantum computation~\cite{Bombin14, Kubica15}.
 
While characterization of excitations and classification of logical operators are intimately related, excitations arising from logical phase gates in the color code have not been studied. In this paper, we study excitations created by transversal phase gates in the color code. Somewhat surprisingly, we find that such excitations can be characterized by bosonic symmetry-protected topological (SPT) phases, which have been actively discussed in condensed matter physics community~\cite{Chen11, Chen11b, Lu12, schuch11, Levin12, Chen13, Vishwanath13, Wen13}. Formally, the system with SPT order has certain on-site symmetry $G$ and its non-degenerate ground state does not break any of the symmetries. Studies of SPT phases have provided better understanding of various quantum phases of matter, including topological insulators, the topological gauge theory and gauge/gravitational anomalies. However, studies of SPT phases have yet to find interesting applications in quantum information science except for few instances~\cite{Cai10, Else12, Else12b}. 

Our first result concerns an observation that $(d-1)$-dimensional excitations, created by transversal $R_{d}$ phase operators in the $d$-dimensional color code, can be characterized by $(d-1)$-dimensional bosonic SPT phases with $(\mathbb{Z}_{2})^{\otimes d}$ symmetry. Namely, by writing the emerging wavefunction as a superposition of excited eigenstates, we find that its expression is identical to a fixed-point wavefunction of a non-trivial SPT phase.  The on-site symmetries emerge from parity conservation of electric charges in the color code. While these SPT excitations are localized on $(d-1)$-dimensional boundaries, their creations require quantum operations acting on all qubits inside the boundaries, reflecting the non-triviality of SPT wavefunctions. 

Although these SPT excitations are pseudo-excitations (not being eigenstates), they may emerge as gapped domain walls in the Heisenberg picture. Our second result concerns an observation that application of transversal logical gates to a part of the system creates a gapped domain wall. In the two-dimensional color code, there is a one-to-one correspondence between transparent domain walls and two-dimensional logical gates with non-trivial automorphism of anyons. In the three-dimensional color code, gapped domain walls, created by $R_{3}$ operators, transform a loop-like magnetic flux into a composite of a magnetic flux and a loop-like SPT excitation, revealing rich possibilities of boundaries in higher-dimensional TQFTs. 

We also study braiding statistics of loop-like excitations in the three-dimensional color code. Our third result concerns an observation that loop-like magnetic fluxes and loop-like SPT excitations exhibit non-trivial three-loop braiding statistics. Namely, if a magnetic flux and a loop-like SPT excitation are braided while both of them are pierced through a magnetic flux, the resulting statistical phase is non-trivial. The non-trivial three-loop braiding statistics results from the fact that the three-dimensional color code admits a fault-tolerantly implementable logical gate from the third-level of the Clifford hierarchy. We also find that excitations, which may condense on the domain wall, exhibit trivial three-loop braiding statistics, implying that domain walls and boundaries in three-dimensional TQFTs may be classified by three-loop braiding statistics of magnetic fluxes and loop-like SPT excitations. 

While the discussion in this paper is limited to a very specific model of topological quantum codes, we believe that our characterization is more generically applicable. Namely, we anticipate that in a large class of topologically ordered systems, pseudo-excitations resulting from fault-tolerantly implementable logical gates can be characterized by SPT wavefunctions. We further expect that these SPT excitations possess non-trivial multi-excitation braiding statistics and provide useful insight into classification of gapped boundaries. We thus view results in this paper as a stepping stone toward establishing the connection between characterizations of gapped domain walls, fault-tolerant logical gates and braiding statistics of SPT excitations.

This paper is organized as follows. In section~\ref{sec:2dim_op}, we describe string and membrane operators in the two-dimensional color code. In section~\ref{sec:2dim_ham}, we show that a loop-like excitation in the two-dimensional color code can be characterized by an SPT wavefunction with $\mathbb{Z}_{2}\otimes \mathbb{Z}_{2}$ symmetry. In section~\ref{sec:2dim_wall}, we argue that SPT excitations can be physically realized as transparent domain walls in the Heisenberg picture. In section~\ref{sec:3dim_op}. we describe string, membrane and volume operators in the three-dimensional color code. In section~\ref{sec:3dim_ham}, we show that a membrane-like excitation in the three-dimensional color code can be characterized by an SPT wavefunction with $\mathbb{Z}_{2}\otimes \mathbb{Z}_{2}\otimes \mathbb{Z}_{2}$ symmetry. In section~\ref{sec:3dim_wall}, we study the transparent domain wall in the three-dimensional color code. We also study the three-loop braiding statistics of magnetic fluxes and loop-like SPT excitations.

\section{Membrane-like operators in two-dimensional color code}\label{sec:2dim_op}

\begin{figure}[htb!]
\centering
\includegraphics[width=0.80\linewidth]{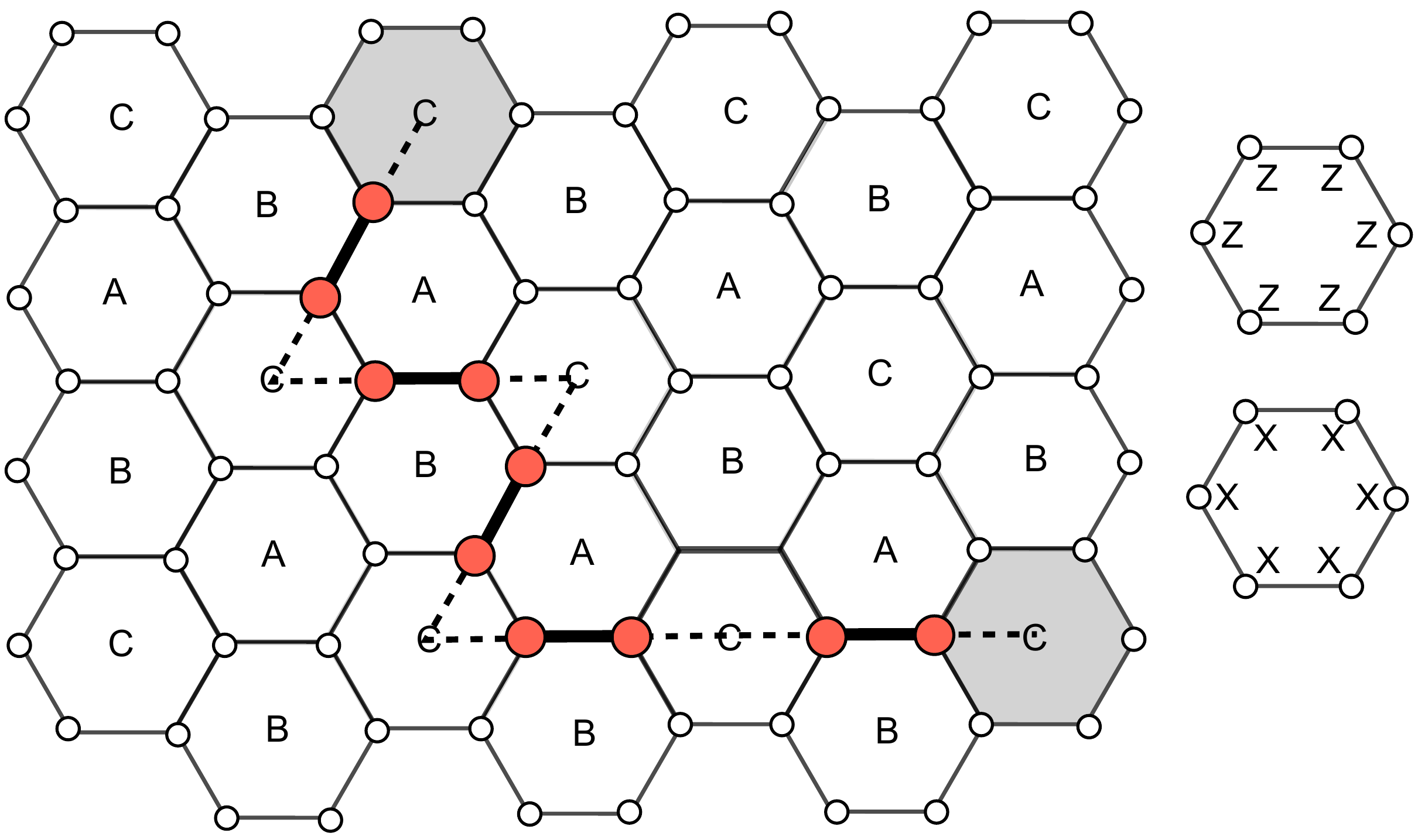}
\caption{The two-dimensional topological color code. The Hamiltonian $H$ is a sum of $X$-type and $Z$-type plaquette terms on every plaquette. An open line $\gamma^{AB}$, consisting of thick edges of color $AB$, defines a string-like operator which creates a pair of anyonic excitations on shaded plaquettes of color $C$.
} 
\label{fig_color_code}
\end{figure}

We begin by considering the two-dimensional color code defined on a three-valent and three-colorable lattice $\Lambda$ where qubits live on vertices. Colors are denoted by $A,B,C$. An example of such a lattice is a hexagonal lattice shown in Fig.~\ref{fig_color_code} where plaquettes are colored in $A,B,C$ such that neighboring plaquettes do not have the same color. The Hamiltonian is given by
\begin{align}
H = - \sum_{P}S^{(X)}_{P}-\sum_{P}S^{(Z)}_{P}
\end{align}
where $P$ represents a plaquette, and $S^{(X)}_{P},S^{(Z)}_{P}$ are tensor products of Pauli-$X,Z$ operators acting on all qubits on a plaquette $P$. Interaction terms $S^{(X)}_{P},S^{(Z)}_{P'}$ commute with each other for all $P,P'$, and thus the system is a stabilizer Hamiltonian. Namely, a ground state $|\psi\rangle$ satisfies stabilizer conditions $S^{(X)}_{P}|\psi\rangle=S^{(Z)}_{P}|\psi\rangle=+|\psi\rangle$ for all $P$. 

Anyonic excitations in two-dimensional topologically ordered spin systems are characterized by string operators. To construct them in the color code, we assign color labels $AB,BC,CA$ to edges of the lattice $\Lambda$ depending on color labels of two adjacent plaquettes. Consider a set of edges of color $AB$ which form a one-dimensional line $\gamma^{AB}$ (Fig.~\ref{fig_color_code}). We define 
\begin{align}
\overline{X^{AB}}|_{\gamma^{AB}} := \bigotimes_{j \in \gamma^{AB}} X_{j}, \quad\overline{Z^{AB}}|_{\gamma^{AB}}  := \bigotimes_{j \in \gamma^{AB}} Z_{j}.
\end{align}
If $\gamma^{AB}$ is an open line, they commute with all the interaction terms except stabilizers on plaquettes of color $C$ at the endpoints of $\gamma_{AB}$. Thus, applications of $\overline{X^{AB}}|_{\gamma^{AB}}$ and $\overline{Z^{AB}}|_{\gamma^{AB}}$ create magnetic fluxes $m_{C}$ and electric charges $e_{C}$ respectively. The correspondence between anyon labels and string operators may be represented as follows:
\begin{align}
\overline{X^{AB}}|_{\gamma^{AB}} \leadsto m_{C}, \quad \overline{Z^{AB}}|_{\gamma^{AB}} \leadsto e_{C}.
\end{align}
Similarly string operators can be constructed from open lines $\gamma^{BC},\gamma^{CA}$, consisting of edges of color $BC,CA$, which lead to 
\begin{equation}
\begin{split}
&\overline{X^{BC}}|_{\gamma_{BC}} \leadsto m_{A}, \quad \overline{Z^{BC}}|_{\gamma_{BC}} \leadsto e_{A},\\
&\overline{X^{CA}}|_{\gamma_{CA}} \leadsto m_{B}, \quad \overline{Z^{CA}}|_{\gamma_{CA}} \leadsto e_{B}.
\end{split}
\end{equation}
There are important subtleties in the above characterization of anyonic excitations. First, anyonic excitations with three different color labels are not independent from each other since applications of Pauli $X,Z$ operator on a single qubit create composites of anyons $m_{A}m_{B}m_{C},e_{A}e_{B}e_{C}$ respectively. In other words, the following fusion channels exist:
\begin{align}
m_{A}\times m_{B}\times m_{C}=1,\quad e_{A}\times e_{B}\times e_{C}=1.
\end{align}
Second, an electric charge $e_{A}$ exhibits the non-trivial braiding statistics with $m_{B}$, but not with $m_{A}$. This is because $\overline{X^{BC}}|_{\gamma_{BC}}$ and $\overline{Z^{BC}}|_{\gamma_{BC}'}$ always commute with each other for any choice of $\gamma_{BC}$ and $\gamma_{BC}'$. To fully capture the braiding statistics in the two-dimensional color code, it is convenient to construct an isomorphism between anyons of the color code and those of the toric code. Let $e_{1},m_{1}$ and $e_{2},m_{2}$ be anyons in two decoupled (\emph{i.e.} non-interacting) copies of the toric code. Then the following correspondence is an isomorphism which preserves braiding and fusion rules:
\begin{align}
m_{A} \leftrightarrow m_{1}, \quad m_{B} \leftrightarrow m_{2},\quad
e_{A} \leftrightarrow  e_{2}, \quad e_{B} \leftrightarrow  e_{1}.
\end{align} 
In fact, it is known that, on a closed manifold, the two-dimensional color code is equivalent to two decoupled copies of the toric code under a local unitary transformation~\cite{Beni10b, Bombin14b, Kubica15b}. In other words, they belong to the same topological phase~\cite{Chen10}.

The two-dimensional color code possesses not only string-like operators, but also transversal membrane (two-dimensional) operators. Let $\mathcal{H}$ be a Hadamard operator which exchanges Pauli $X$ and $Z$ operators:
\begin{align}
\mathcal{H}X\mathcal{H}^{\dagger}=Z,\quad \mathcal{H}Z\mathcal{H}^{\dagger}=X.
\end{align}
The Hamiltonian $H$ is symmetric under transversal conjugation by Hadamard operators
\begin{align}
\overline{\mathcal{H}} H \overline{\mathcal{H}}^{\dagger} = H,\quad \overline{\mathcal{H}}:=\bigotimes_{j}\mathcal{H}_{j}.
\end{align}
Since $\overline{\mathcal{H}}$ transforms $X$-type string operators into $Z$-type string operators and vise versa, it exchanges electric charges $e$ and magnetic fluxes $m$:
\begin{align}
e_{A} \rightarrow m_{A}, \quad  e_{B} \rightarrow m_{B}, \quad  m_{A} \rightarrow e_{A},\quad m_{B} \rightarrow e_{B}.
\end{align}
The color code admits another interesting transversal menbrane operator. Let us define phase operators, acting on a qubit, by $R(\theta) := \mbox{diag}(1,e^{i\theta})$. Of particular importance is the so-called $R_{m}$ phase operator
\begin{align}
R_{m}:=\mbox{diag}(1,\exp(i\pi/2^{m-1})).
\end{align}
The $R_{2}$ operator exchanges Pauli $X$ and $Y$ operators:
\begin{align}
R_{2}XR_{2}^{\dagger}=Y,\quad R_{2}YR_{2}^{\dagger}=-X.
\end{align}
Let $\Pi$ be a projector onto the ground state space of the color code Hamiltonian. Recall that the lattice $\Lambda$ is bipartite and qubits can be split into two complementary sets $\mathcal{T}$ and $\mathcal{T}^{c}$. Let us define the following transversal (two-dimensional) phase operator
\begin{align}
\overline{R_{2}} := \bigotimes_{j \in \mathcal{T}} R_{2}|_{j} \bigotimes_{j \in \mathcal{T}^{c}} (R_{2}|_{j})^{-1}.
\end{align}
Then the ground state space is invariant under transversal application of $R_{2}$ operators: $\overline{R_{2}} \Pi= \Pi\overline{R_{2}}\Pi$.
This two-dimensional membrane operator implements the following exchanges of anyon labels (an automorphism):
\begin{align}
e_{A} \rightarrow e_{A}, \  e_{B} \rightarrow e_{B}, \  m_{A} \rightarrow m_{A}e_{A},\ m_{B} \rightarrow m_{B}e_{B}.
\end{align}
In general, in two-dimensional topologically ordered spin systems described by TQFTs, transversal membrane-like (two-dimensional) operators may induce an automorphism of anyon labels which preserves braiding and fusion rules (\emph{i.e.} monoidal centers of categories which define $(2+1)$-dimensional TQFTs)~\cite{Kitaev12, Beverland14}. 
 
We conclude this section by recalling quantum information theoretical motivations to study transversal membrane-like operators in two-dimensional topologically ordered spin systems. 
In quantum information science, one hopes to perform quantum information processing tasks in a protected codeword space of some quantum error-correcting code. The gapped ground state space of topologically ordered systems is an idealistic platform for such purposes. But how do we perform quantum computations inside the protected subspace? Ideally one hopes to perform logical operations in a way which does not make local errors propagate to other spins. Namely, one hopes to perform logical operations by transversal unitary gates acting on each spin as a tensor product. Thus, it is important to classify transversally implementable logical gates in quantum error-correcting codes~\cite{Eastin09, Bravyi13b, Pastawski15, Beverland14}. 
Transversal membrane operators, such as $\overline{\mathcal{H}}$ and $\overline{R_{2}}$ in the topological color code, are examples of fault-tolerantly implementable logical gates as they may have non-trivial action on the ground state space (if it is degenerate). Our goal is to characterize excitations arising from fault-tolerantly implementable logical operators, and the present paper is dedicated to studies of those in the topological color code.

\section{SPT excitations in two-dimensional topological color code}\label{sec:2dim_ham}

In this section, we study loop-like excitations created by parts of a membrane phase operator $\overline{R_{2}}$ in the two-dimensional color code and show that they are characterized by a wavefunction of a one-dimensional bosonic SPT phase with $\mathbb{Z}_{2}\otimes \mathbb{Z}_{2}$ symmetry. Our finding reveals that these loop-like excitations in the color code can be viewed as a path integral formulation of an SPT wavefunction, leading to a physically insightful proof that such a wavefunction cannot be prepared by symmetry-protected local unitary transformations. We note that the circuit depth of preparing SPT wavefunctions was previously studied by using different approaches~\cite{Marvian13, Huang14}. 

In this section, for simplicity of discussion, we assume that the lattice $\Lambda$ is supported on a sphere so that the system has a unique ground state $|\psi_{gs}\rangle$. 

\subsection{Loop-like excitation from membrane operator}

To begin, let $R(\theta)$ be a phase operator $R(\theta):=\mbox{diag}(1,e^{i\theta})$. An application of $R(\theta)$ on a qubit at a vertex $v$ creates an excited wavefunction:
\begin{align}
|\psi(\theta)\rangle={R(\theta)}|_{v}\cdot |\psi_{gs}\rangle. 
\end{align}
Since $R(\theta)$ is diagonal in the computational basis, it creates excitations which are associated with $X$-type stabilizers on three neighboring plaquettes $P_{1},P_{2},P_{3}$ of three different colors. We would like to characterize this wavefunction in the \emph{excitation basis}:
\begin{align}
|\psi(\theta)\rangle \mapsto \sum_{p_{1},p_{2},p_{3}} \lambda_{p_{1},p_{2},p_{3}} |\tilde{p_{1}},\tilde{p_{2}},\tilde{p_{3}}\rangle
\end{align}
where $ \tilde{p_{1}},\tilde{p_{2}},\tilde{p_{3}}=0,1$ and $|\tilde{p_{1}},\tilde{p_{2}},\tilde{p_{3}}\rangle $ represents an eigenstate of the Hamiltonian $H$ with 
\begin{equation}
\begin{split}
S^{(X)}_{P_{1}}|\tilde{p_{1}},\tilde{p_{2}},\tilde{p_{3}}\rangle &= (1-2p_{1})|\tilde{p_{1}},\tilde{p_{2}},\tilde{p_{3}}\rangle \\ 
S^{(X)}_{P_{2}}|\tilde{p_{1}},\tilde{p_{2}},\tilde{p_{3}}\rangle &= (1-2p_{2})|\tilde{p_{1}},\tilde{p_{2}},\tilde{p_{3}}\rangle\\
S^{(X)}_{P_{3}}|\tilde{p_{1}},\tilde{p_{2}},\tilde{p_{3}}\rangle &= (1-2p_{3})|\tilde{p_{1}},\tilde{p_{2}},\tilde{p_{3}}\rangle. 
\end{split}
\end{equation}
In other worlds, $|p_{1},p_{2},p_{3} \rangle $ records the presence or absence of electric charges at $P_{1},P_{2},P_{3}$ while there is no other excitations in the system. One finds
\begin{align}
|\psi(\theta)\rangle \mapsto \cos(\theta/2)|\tilde{0},\tilde{0},\tilde{0}\rangle + i\sin(\theta/2)|\tilde{1},\tilde{1},\tilde{1}\rangle. \label{eq:transformation}
\end{align}
So the phase operator $R(\theta)$ corresponds to
\begin{align}
R(\theta)\mapsto \tilde{R}(\theta) = \exp\left(i\frac{\theta}{2} \tilde{X}_{P_{1}}\tilde{X}_{P_{2}}\tilde{X}_{P_{3}}\right)
\end{align}
where $\tilde{X}_{P_{1}},\tilde{X}_{P_{2}},\tilde{X}_{P_{3}}$ act on $|\tilde{0}\rangle$ and $|\tilde{1}\rangle$ as Pauli $X$ operators. Here we used ``$\mapsto$'' to denote the map from real physical systems to the excitation basis. 

Now consider a subset of qubits $V$ and a restriction of the membrane phase operator $\overline{R_{2}}$ onto $V$, denoted by $R_{2}|_{V}$:
\begin{align}
\overline{R_{2}} := \bigotimes_{j \in V\cap \mathcal{T}} R_{2}|_{j} \bigotimes_{j \in V\cap \mathcal{T}^{c}} (R_{2}|_{j})^{-1}.
\end{align}
Consider an excited wavefunction 
\begin{align}
|\psi_{V}\rangle:=R_{2}|_{V}\cdot |\psi_{gs}\rangle.
\end{align}
We hope to represent $|\psi_{V}\rangle$ in the excitation basis. First let us be more precise with a definition of the excitation basis states. Let $\mathcal{H}_{\textrm{no-flux}}$ be the fluxless subspace of the entire Hilbert space where $|\psi\rangle \in \mathcal{H}_{\textrm{no-flux}}$ satisfies $S^{(Z)}_{P}|\psi\rangle = + |\psi\rangle$ for all $P$. Let $n_{0}$ be the total number of plaquettes on the lattice $\Lambda$. We define the excitation basis states $|\tilde{p_{1}},\ldots,\tilde{p_{n_{0}}}\rangle \in \mathcal{H}_{\textrm{no-flux}}$ by
\begin{align}
S^{(X)}_{P_{j}}|\tilde{p_{1}},\ldots,\tilde{p_{n_{0}}}\rangle=(1-2p_{j})|\tilde{p_{1}},\ldots,\tilde{p_{n_{0}}}\rangle
\end{align}
for $j=1,\ldots,n_{0}$. More explicitly, we define them by
\begin{align}
|\tilde{p_{1}},\ldots,\tilde{p_{n_{0}}}\rangle := \frac{1}{2^{n_{0}/2}}\prod_{j=1}^{n_{0}}\big(1+ (-1)^{\tilde{p_{j}}} S^{(X)}_{P_{j}}\big)|0\cdots\rangle. \label{eq:definition}
\end{align}
Importantly, not all the basis states are physically allowed since there are certain $\mathbb{Z}_{2}$ constraints on values of $\tilde{p_{j}}$. Observe that
\begin{align}
\prod_{P \in \mathcal{A}} S^{(X)}_{P}=\prod_{P\in \mathcal{B}} S^{(X)}_{P}=\prod_{P\in \mathcal{C}} S^{(X)}_{P} = \bigotimes_{\forall j} X_{j}
\end{align}
where $\mathcal{A}, \mathcal{B}, \mathcal{C}$ represent sets of plaquettes of color $A,B,C$ respectively. Let $N_{\mathcal{A}}$, $N_{\mathcal{B}}$, $N_{\mathcal{C}}$ be the total number of excitations on plaquettes of color $A,B,C$ respectively. Then,
\begin{align}
N_{\mathcal{A}} = N_{\mathcal{B}}= N_{\mathcal{C}} \qquad \mbox{(mod $2$)}. \label{eq:constraint}
\end{align} 
In other words, an excitation basis state $|\tilde{p_{1}},\ldots,\tilde{p_{n_{0}}}\rangle$ is physically allowed if and only if $\sum_{j : P_{j}\in \mathcal{A}}\tilde{p_{j}}=\sum_{j : P_{j}\in \mathcal{B}}\tilde{p_{j}}=\sum_{j : P_{j}\in \mathcal{C}}\tilde{p_{j}}$ modulo $2$. (Indeed, if $|\tilde{p_{1}},\ldots,\tilde{p_{n_{0}}}\rangle$ does not satisfy this condition, then the righthand side of Eq.~(\ref{eq:definition}) becomes zero). Because the system is not degenerate, basis states $|\tilde{p_{1}},\ldots,\tilde{p_{n_{0}}}\rangle$ with Eq.~(\ref{eq:constraint}) span the fluxless subspace $\mathcal{H}_{\textrm{no-flux}}$ completely. Thus, the excited wavefunction can be characterize as follows
\begin{align}
|\psi_{V}\rangle \mapsto \sum_{\tilde{p_{1}},\ldots,\tilde{p_{n_{0}}}} \lambda_{\tilde{p_{1}},\ldots,\tilde{p_{n_{0}}}} |\tilde{p_{1}},\ldots,\tilde{p_{n_{0}}}\rangle
\end{align}
in the excitation basis where $\lambda_{\tilde{p_{1}},\ldots,\tilde{p_{n_{0}}}}$ is a complex number with proper normalization and $\tilde{p_{1}},\ldots,\tilde{p_{n_{0}}}$ satisfy Eq.~(\ref{eq:constraint}). 

\begin{figure}[htb!]
\centering
\includegraphics[width=0.90\linewidth]{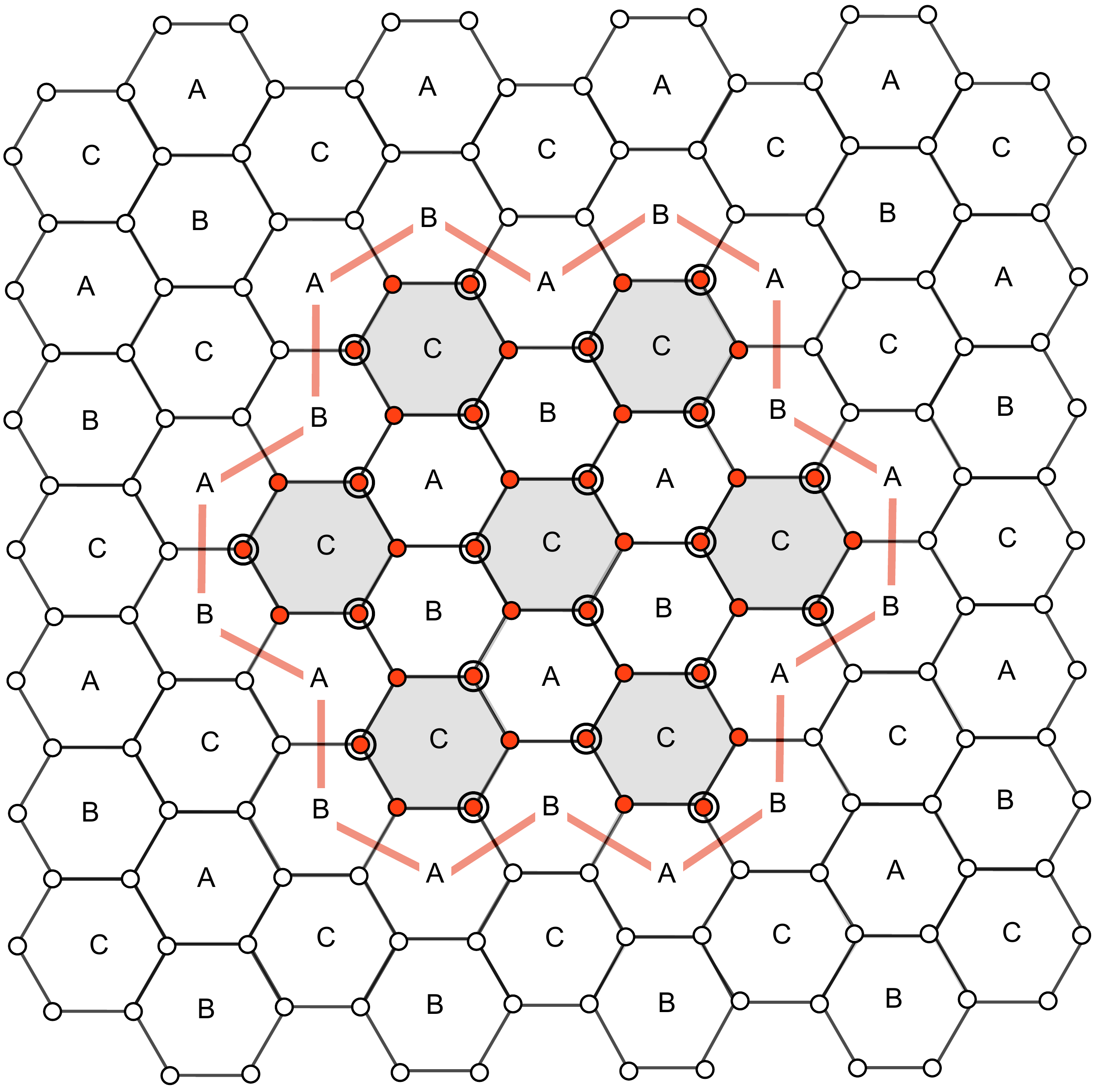}
\caption{A loop-like excitation created by $R_{2}$ phase operators. Filled dots represent qubits in $V$ and plaquettes crossed by a closed loop around $V$ form $\partial V$. The region $V$ is constructed from a set of plaquettes of color $C$. One applies $R_{2}$ operators on filled circles and $(R_{2})^{-1}$ on filled double circles. 
} 
\label{fig_boundary}
\end{figure}

Let us then study a loop-like excitation created by a part of the membrane-like $\overline{R_{2}}$ operator. Consider a set of plaquettes of color $C$ which form a contractable connected region of qubits $V$ with a single boundary (see Fig.~\ref{fig_boundary}). Consider an excited wavefunction $|\psi_{V}\rangle=\overline{R_{2}}|_{V}\cdot |\psi_{gs}\rangle$. Since $\overline{R_{2}}|\psi_{gs}\rangle=|\psi_{gs}\rangle$, the phase operator $\overline{R_{2}}|_{V}$ creates excitations only around the boundary of $V$. A key observation is that $\overline{R_{2}}|_{V}$ creates excitations \emph{only on plaquettes of color $A,B$} because $V$ can be viewed as a set of plaquettes of color $C$. Namely, let $\partial V$ be a set of plaquettes of color $A,B$ on the boundary (Fig.~\ref{fig_boundary}). Observe that, on the boundary $\partial V$, plaquettes of color $A$ and $B$ appear in an alternating way. We denote boundary plaquettes by $A_{1},B_{1},\ldots,A_{n},B_{n}$ where $2n$ is the total number of boundary plaquettes. Then the excitation wavefunction can be written as 
\begin{align}
|\psi_{V}\rangle \mapsto |\phi^{(\textrm{ex})}_{\partial V}\rangle \otimes |\tilde{0},\ldots,\tilde{0}\rangle
\end{align}
where $|\phi^{(\textrm{ex})}_{\partial V}\rangle = \sum_{\vec{p}} \lambda_{\vec{p}} |\tilde{p_{A_{1}}},\tilde{p_{B_{1}}}, \ldots,\tilde{p_{A_{n}}},\tilde{p_{B_{n}}} \rangle$.
Here, the first part $|\phi^{(\textrm{ex})}_{\partial V}\rangle$  represents excitations on the boundary and the second part $|\tilde{0},\ldots,\tilde{0}\rangle$ represents the rest. 

The next task is to find an expression of $|\phi_{\partial V}^{(\textrm{ex})}\rangle$. From Eq.~(\ref{eq:transformation}), the excited wavefunction is given by
\begin{align}
|\psi_{V}\rangle \mapsto \prod_{\langle i,j,k \rangle\in V} \exp\left(\pm  i \frac{\pi}{4} \tilde{X}_{i}\tilde{X}_{j}\tilde{X}_{k} \right)\cdot |\tilde{0}\rangle^{\otimes n_{0}}
\end{align}
where $\pm$ in the product corresponds to $R_{2}$ and $(R_{2})^{-1}$ in the phase operator $\overline{R_{2}}$ respectively. Here $\langle i,j,k \rangle$ represents a vertex shared by three neighboring plaquettes $i,j,k$. After careful calculations, one can find that the boundary wavefunction, in the excitation basis, is given by
\begin{align}
|\phi^{(\textrm{ex})}_{\partial V}\rangle = \tilde{U}_{2}|_{\partial V} \cdot |\tilde{0}\rangle^{\otimes 2n}.
\end{align}
where 
\begin{align}
\tilde{U}_{2}|_{\partial V} :=\prod_{j=1}^{n}\exp\left(i\frac{\pi}{4}\tilde{X}_{A_{j}}\tilde{X}_{B_{j}} \right)\cdot \exp\left(-i\frac{\pi}{4}\tilde{X}_{B_{j}}\tilde{X}_{A_{j+1}}\right).
\end{align}
The boundary wavefunction $|\phi^{(\textrm{ex})}_{\partial V}\rangle$ can be viewed as a one-dimensional system of $2n$ qubits supported on a closed loop. It is worth finding the Hamiltonian which has the boundary wavefunction $|\phi^{(\textrm{ex})}_{\partial V}\rangle$ as a unique ground state. Recall that $\tilde{Z}_{P}|\tilde{0}\rangle^{\otimes n_{0}}=|\tilde{0}\rangle^{\otimes n_{0}}$ for all plaquettes $P$. Interaction terms in the Hamiltonian are then obtained by considering $U_{2}|_{\partial V} \tilde{Z}_{P}U_{2}|_{\partial V}^{\dagger}$. One finds
\begin{align}
\tilde{H}_{\partial V} = - \sum_{j} \tilde{X}_{A_{j-1}}\tilde{Z}_{B_{j-1}}\tilde{X}_{A_{j}} - \sum_{j}\tilde{X}_{B_{j-1}} \tilde{Z}_{A_{j}}\tilde{X}_{B_{j}}.
\end{align}
One can verify that $|\phi^{(\textrm{ex})}_{\partial V}\rangle$ is the unique gapped ground state of the Hamiltonian $\tilde{H}_{\partial V}$. This Hamiltonian for the boundary wavefunction is identical to that of the so-called cluster state up to transversal application of the Hadamard operators to each and every qubit along the boundary: 
\begin{align}
H_{\textrm{cluster}} = - \sum_{j=1}^{2n} Z_{j-1}X_{j}Z_{j+1}  = \mathcal{H}^{\otimes 2n} \tilde{H}_{\partial V} \mathcal{H}^{\otimes 2n}
\label{eq:cluster_ham}
\end{align}
The ground state (the cluster state) is specified by 
\begin{align}
Z_{j-1}X_{j}Z_{j+1}|\psi\rangle =|\psi\rangle \qquad \mbox{for all $j$.}
\end{align}
In quantum information science community, the cluster state is known as an important resource state for realizing the measurement-based quantum computation scheme~\cite{Raussendorf03}. 

\subsection{SPT excitation with $\mathbb{Z}_{2}\otimes \mathbb{Z}_{2}$ symmetry}

The cluster state is perhaps the simplest example of one-dimensional bosonic SPT phases with $\mathbb{Z}_{2}\otimes \mathbb{Z}_{2}$ symmetry. Let us define a pair of $\mathbb{Z}_{2}$ on-site symmetry operators as follows:
\begin{align}
\mathcal{S}_{A} := \bigotimes_{j=1}^{n} X_{2j} \qquad \mathcal{S}_{B} := \bigotimes_{j=1}^{n} X_{2j-1}.\label{eq:2d_symmetry_hadamard}
\end{align}
By on-site, we mean that symmetry operators are transversal. The cluster state $|\psi_{\textrm{cluster}}\rangle$ is the unique ground state of the Hamiltonian (Eq.~(\ref{eq:cluster_ham})) and is symmetric under $\mathcal{S}_{A}$ and $\mathcal{S}_{B}$:
\begin{align}
\mathcal{S}_{A}|\psi_{\textrm{cluster}}\rangle = \mathcal{S}_{B} |\psi_{\textrm{cluster}}\rangle =|\psi_{\textrm{cluster}}\rangle.
\end{align}
This can be verified by noticing $\prod_{j=even}Z_{j-1}X_{j}Z_{j+1}=\mathcal{S}_{A}$ and $\prod_{j=odd}Z_{j-1}X_{j}Z_{j+1}=\mathcal{S}_{B}$, and $Z_{j-1}X_{j}Z_{j+1}|\psi_{\textrm{cluster}}\rangle =|\psi_{\textrm{cluster}}\rangle$. Let $|\psi_{\textrm{trivial}}\rangle = |+\rangle^{\otimes 2n}$ be a trivial product state with $\mathbb{Z}_{2}\otimes \mathbb{Z}_{2}$ symmetry where $|+\rangle := \frac{1}{\sqrt{2}}(|0\rangle + |1\rangle)$. Two symmetric wavefunctions $|\psi_{\textrm{cluster}}\rangle$ and $|\psi_{\textrm{trivial}}\rangle$ are connected by a local unitary transformation and belong to the same quantum phase in the absence of symmetries. Indeed, one has
\begin{align}
|\psi_{\textrm{cluster}}\rangle = \prod_{j=1}^{2n} \exp\left( i\frac{\pi}{4} Z_{j}Z_{j+1} \right) |\psi_{\textrm{trivial}}\rangle. \label{eq:cluster}
\end{align}
On the other hand, in the presence of $\mathbb{Z}_{2}\otimes \mathbb{Z}_{2}$ symmetry, they belong to different SPT phases. Namely, there is no symmetry-protected local unitary transformation $U$ such that $[U,\mathcal{S}_{A}]=[U,\mathcal{S}_{B}]=0$ and $|\psi_{\textrm{cluster}}\rangle = U|\psi_{\textrm{trivial}}\rangle$. For instance, one sees that the unitary transformation in Eq.~(\ref{eq:cluster}) does not commute with $\mathcal{S}_{A}$ or $\mathcal{S}_{B}$. In this sense, a cluster state is a non-trivial SPT wavefunction with $\mathbb{Z}_{2}\otimes \mathbb{Z}_{2}$ symmetry. 

It is interesting to observe that loop-like excitations in the two-dimensional color code are characterized by wavefunctions of a one-dimensional SPT phase. To understand what this observation means, let us further establish the connection between SPT phases and loop-like excitations. Indeed, on-site symmetry of the boundary wavefunction $|\phi^{(\textrm{ex})}_{\partial V}\rangle$ naturally emerges from parity conservation on the number of anyonic excitations in the topological color code. Since $R_{2}$ phase operators are applied on qubits supported on plaquettes of color $C$, there is no excitation associated with plaquettes of color $C$. This implies 
\begin{align}
\mathcal{N}_{A}=\mathcal{N}_{B}=0 \qquad \mbox{(mod $2$)}
\end{align}
which leads to the following on-site symmetries in the excitation basis
\begin{align}
\tilde{\mathcal{S}}_{A}=\prod_{j} \tilde{Z}_{A_{j}} \qquad \tilde{\mathcal{S}}_{B}=\prod_{j} \tilde{Z}_{B_{j}}.\label{eq:2d_symmetry}
\end{align}
After transversal Hadamard transformation, these symmetry operators are identical to those in Eq.~(\ref{eq:2d_symmetry_hadamard}). As such, excitations arising in the two-dimensional color code are natural platforms for constructing wavefunctions with  $\mathbb{Z}_{2}\otimes \mathbb{Z}_{2}$ symmetry. 

\begin{figure}[htb!]
\centering
\includegraphics[width=0.55\linewidth]{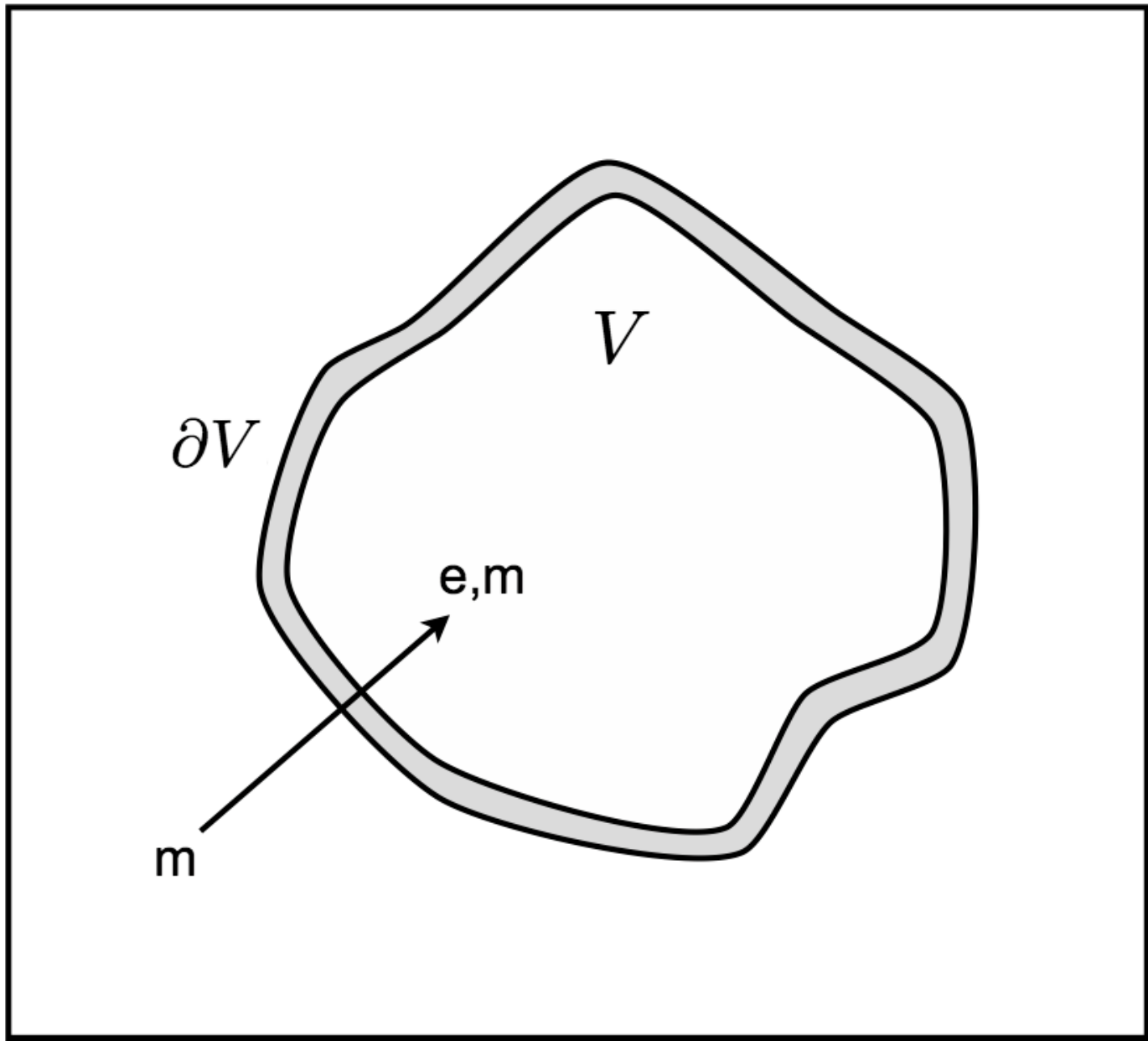}
\caption{An SPT excitation localized on the boundary $\partial V$. Creation of this loop-like excitation on $\partial V$ requires operators acting on all qubits in $V$. If a magnetic flux $m$ crosses an SPT excitation, it gets transformed into a composite of a magnetic flux and an electric charge $e$. 
} 
\label{fig_SPT_excitation}
\end{figure}

The remaining question then is why the boundary wavefunction corresponds to a non-trivial SPT phase. The key observation is that, while these loop-like excitations are localized along the boundary $\partial V$, they cannot be created by a local unitary transformation acting on physical qubits in the neighborhood of $\partial V$. To see this, let us create a pair of magnetic fluxes which are located outside of $V$ and move one of them inside $V$, crossing the SPT excitation (Fig.~\ref{fig_SPT_excitation}). Since $R_{2}$ phase operators exchange Pauli $X$ and $Y$ operators, the magnetic flux $m$ will be transformed into a composite of an electric charge $e$ and a magnetic flux $m$ upon crossing the SPT excitation. Let $\ell$ be a string operator corresponding to the propagation of a magnetic flux into $V$ in the absence of an SPT excitation. (It is a tensor product of Pauli $X$ operators). Suppose that there exists a local unitary $U$ which creates an SPT excitation by acting only on qubits in the neighborhood of $\partial V$. Then $U\ell U^{\dagger}$ differs from $\ell$ only at the intersection with the boundary. This implies that a magnetic flux remains to be a magnetic flux inside $V$, leading to a contradiction. Thus, to create a loop-like SPT excitation, one needs to apply a local unitary transformation on all qubits inside $V$ (or all qubits in the complement of $V$). 

This argument enables us to show that two symmetric wavefunctions $|\psi_{\textrm{cluster}}\rangle$ and $|\psi_{\textrm{trivial}}\rangle$ belong to different SPT phases. This is because symmetry-protected local unitary operators in the excitation basis have some corresponding local unitary operators in the two-dimensional color code. Suppose that there exists a symmetry-protected local unitary transformation $\tilde{U}$ such that $|\psi_{\textrm{cluster}}\rangle = \tilde{U}|\psi_{\textrm{trivial}}\rangle$, which can be generically written as 
\begin{align}
\tilde{U} = \mathcal{T}\left[ \int_{0}^{1} \exp( -i H_{sp}(t) )dt\right] 
\end{align}
where $[H_{sp}(t),\tilde{\mathcal{S}}_{A}]=[H_{sp}(t),\tilde{\mathcal{S}}_{B}]=0$ and $H_{sp}(t)$ is geometrically local and consists only of terms with bounded norms. Here $\mathcal{T}$ represents the time-ordering. Note $H_{sp}(t)$ can be written as a sum of symmetry-protected local terms such as $X_{A_{i}}X_{A_{j}}$, $X_{B_{i}}X_{B_{j}}$, $Z_{A_{i}}$, $Z_{B_{i}}$ and their products. These symmetry-protected local unitary operators correspond to some local operators in the topological color code. Namely, $X_{A_{i}}X_{A_{j}}$ and $X_{B_{i}}X_{B_{j}}$ correspond to string operators of color $BC$ and $CA$, consisting of Pauli $Z$ operators, which end at plaquettes $A_{i},A_{j}$ and $B_{i},B_{j}$ respectively. Moreover, $Z_{A_{i}}$ and $Z_{B_{i}}$ correspond to $X$-type plaquette operators of color $A,B$. As such, the existence of a symmetry-protected local unitary $\tilde{U}$ implies the existence of a local unitary transformation $U$ which acts only on physical qubits near the boundary, yet creates a loop-like SPT excitation. This leads to a contradiction. Thus, we can conclude that $|\psi_{\textrm{cluster}}\rangle$ and $|\psi_{\textrm{trivial}}\rangle$ belong to different SPT phases.

We have argued that an excited wavefunction $|\phi^{(\textrm{ex})}_{\partial V}\rangle$ corresponds to a non-trivial SPT phase due to the $\mathbb{Z}_{2}\otimes \mathbb{Z}_{2}$ parity conservation and the non-trivial automorphism of anyons induced by the $\overline{R_{2}}$ phase operator. In general, loop-like excitations may involve plaquettes of three different colors when $V$ is chosen to be an arbitrary connected region of qubits. The excited wavefunction then can be viewed as a one-dimensional system with spins of three different colors which possesses $\mathbb{Z}_{2}\otimes \mathbb{Z}_{2}$ symmetry with respect to $\tilde{\mathcal{S}}_{A}\tilde{\mathcal{S}}_{B}\otimes \tilde{\mathcal{S}}_{A}\tilde{\mathcal{S}}_{C}$ where $\tilde{\mathcal{S}}_{A},\tilde{\mathcal{S}}_{B},\tilde{\mathcal{S}}_{C}$ are tensor products of Pauli $Z$ operators acting on spins of color $A,B,C$ in the excitation basis. As such, our characterization of excitations by SPT phases is valid in these cases too. 

For readers who are familiar with the literature of SPT phases, it may be clear that the characterization of a wavefunction in the excitation basis in the fluxless subspace $\mathcal{H}_{\textrm{no-flux}}$ is essentially equivalent to ``ungauging'' the on-site symmetries, opposite to the procedure of gauging the on-site symmetries of SPT wavefunctions~\cite{Levin12, Hu13}. In this picture of ungauging, the two-dimensional color code with a loop-like excitation serves as a path integral formulation of a one-dimensional SPT wavefunction. Namely, creation of a loop-like SPT excitation via transversal $R_{2}$ operators on the bulk $V$ can be interpreted as a symmetry-protected quantum circuit preparing a non-trivial SPT wavefunction with $\mathbb{Z}_{2}\otimes \mathbb{Z}_{2}$ symmetry. An interesting application of this picture is that, if one considers the two-dimensional color code embedded on a hyperbolic surface, one obtains a MERA (multi-scale entanglement renormalization ansatz) circuit for an SPT wavefunction with $\mathbb{Z}_{2}\otimes \mathbb{Z}_{2}$~\cite{Vidal07}. 

Finally we remark that similar SPT excitations emerge in two decoupled copies of the toric code where the transversal control-$Z$ operator preserves the ground state space. In this setting, a loop-like SPT excitation involves electric charges from two copies of the toric code, each possessing one copy of the $\mathbb{Z}_{2}$ symmetry. This conclusion also follows from the unitary equivalence of the color code and two decoupled copies of the toric code on a closed manifold~\cite{Kubica15b}. 

\section{Gapped domain walls and fault-tolerant logical gates}\label{sec:2dim_wall}

While one-dimensional SPT excitations, created by $R_{2}$ phase operators in the two-dimensional color code, are interesting from a theoretical viewpoint, they do not exist as stable objects since they are superpositions of eigenstate excitations which would decohere immediately. In this section, we argue that SPT excitations are physically realized in a certain way. Namely, we point out that SPT excitations can be viewed as transparent gapped domain walls in the color code in the Heisenberg picture. Our finding also reveals an intriguing relation between classifications of domain walls and fault-tolerantly implementable logical gates. 

We note that boundaries in the two-dimensional quantum double model are discussed in Ref.~\cite{Beigi11}. We also note that domain walls in SPT phases are discussed in Ref.~\cite{Chen14}.

\begin{figure}[htb!]
\centering
\includegraphics[width=0.55\linewidth]{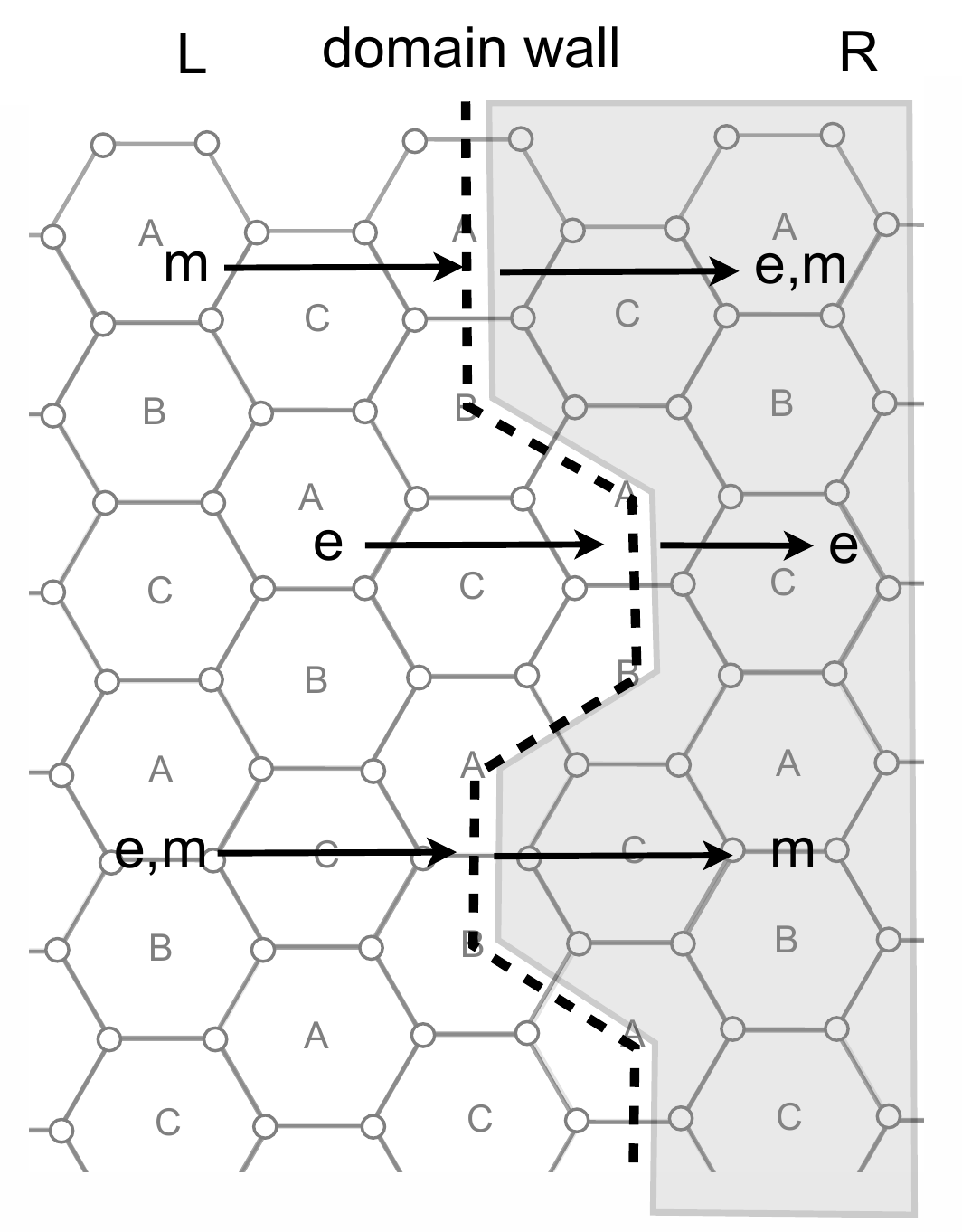}
\caption{Construction of a transparent gapped domain wall in the topological color code. $R_{2}$ operators are applied only to the qubits on the right side of the system (a shared region). Anyons get transformed when crossing the domain wall as depicted in the figure. 
} 
\label{fig_wall}
\end{figure}

\subsection{Transparent domain wall}

The key idea is to transform the Hamiltonian by transversal operators instead of transforming the ground state. To begin, consider the two-dimensional color code supported on the honeycomb lattice (Fig.~\ref{fig_wall}). Recall that the transversal Hadamard operator, $\overline{\mathcal{H}} = \bigotimes_{v} \mathcal{H}_{v}$, preserves the Hamiltonian. Here we split the entire system into two parts, the left part $L$ and the right part $R$. Consider the restriction of the transversal Hadamard operator onto the right part of the system $R$:
\begin{align}
\overline{\mathcal{H}}|_{R} = \bigotimes_{v\in R} \mathcal{H}|_{v}
\end{align}
and the transformed Hamiltonian $\hat{H} = (\overline{\mathcal{H}}|_{R})  H  (\overline{\mathcal{H}}|_{R})^{\dagger}$. Note that the resulting Hamiltonian $\hat{H}$ remains the same as before except on the boundary between $L$ and $R$:
\begin{align}
\hat{H} = H_{L} + H_{R} + H_{LR}
\end{align}
where $H_{L}$ and $H_{R}$ are the same as the corresponding terms in $H$. Since $\overline{\mathcal{H}}|_{R}$ is a unitary transformation, the transformed Hamiltonian $\hat{H}$ remains gapped. Here, $H_{LR}$ can be viewed as a transparent gapped domain wall connecting $H_{L}$ and $H_{R}$, which swaps the electric charge $e$ and magnetic flux $m$ as follows:
\begin{align}
(e_{A}|m_{A}), \quad (m_{A}|e_{A}),\quad
(e_{B}|m_{B}), \quad (m_{B}|e_{B}).
\end{align}
By a transparent wall, we mean that no single anyon from either side of the wall may condense on the wall. Here, $(a|b)$ represents that an anyon $a$ gets transformed into $b$ by crossing the wall from the left to the right. Transparency of the wall imposes that $(a|1)$ or $(1|a)$ for $a\not=1$ is not allowed where $1$ denotes the vacuum. Similarly, the transversal $R_{2}$ phase operator on $R$ creates a gapped domain wall which changes anyon labels as follows: 
\begin{align}
(m_{A} | e_{A}m_{A}), \ (e_{A} | e_{A}),\
(m_{B} | e_{B}m_{B}), \ (e_{B} | e_{B}).
\end{align}
This transparent domain wall corresponds to the loop-like SPT excitation created by $R_{2}$ operators in the Heisenberg picture as shown in Fig.~\ref{fig_wall}. The ground state space of the topological color code is also symmetric under the so-called $T$ transformation:
\begin{equation}
\begin{split}
T \ : \ X \rightarrow Y,\quad Y \rightarrow Z, \quad Z \rightarrow X.
\end{split}
\end{equation}
Transversal application of $T$ operators leads to the following transparent domain wall:
\begin{equation}
\begin{split}
(e_{A} | m_{A}), \ (m_{A} | e_{A}m_{A}), \ (e_{B} | m_{B}), \ (m_{B} | e_{B}m_{B}).
\end{split}
\end{equation}

Clearly, the aforementioned construction of transparent gapped domain walls works for arbitrary transversal membrane-like operators in $(2+1)$-dimensional TQFTs as long as the operators induce non-trivial automorphism among anyon labels. It turns out that the construction works not only for transversal operators but also for any \emph{locality-preserving transformations}~\cite{Pastawski15, Beverland14}. Formally, a locality-preserving unitary transformation $U$ is defined to satisfy the following condition. For an arbitrary unitary operator $\ell$ supported on some region $V$, there exists $\ell'$ supported on $V\cup \partial V$ such that $|U\ell U^{\dagger}- \ell'|\approx 0$ where $\partial V$ is the boundary of $V$ of finite width. Examples of locality-preserving transformations include local unitary transformations with removing and adding ancilla qubits. Also translating all qubits on the lattice by finite sites is a locality-preserving transformation. By applying such a transformation with non-trivial automorphism of anyon labels on the half of the system, one can create a transparent gapped domain wall which changes anyon labels according to the automorphism associated with the membrane-like operator. 

To give a concrete yet non-trivial example of locality-preserving transformation, consider the two-dimensional toric code supported on a square lattice on a torus: $H_{toric}=- \sum_{v}A_{v} - \sum_{p}B_{p}$. Observe that transversal application of Hadamard operators, followed by shifting all the lattice sites in a diagonal direction, leaves the Hamiltonian invariant. This transformation is clearly locality-preserving and swaps electric charge $e$ and magnetic flux $m$. If one applies this transformation partially on one side of the system, one is able to create a domain wall which exchanges $e$ and $m$ upon crossing the wall. (Equivalently, one may consider the Wen's formulation of the toric code where unit translations lead to exchange of $e$ and $m$~\cite{Wen02}). We note that this domain wall in the toric code was previously constructed by Bombin~\cite{Bombin10}.

\subsection{Membrane operator and domain wall}

We have seen that membrane operators associated with non-trivial automorphisms among anyon labels lead to transparent gapped domain walls in $(2+1)$-dimensional TQFTs. An interesting question is whether the presence of transparent gapped domain walls implies membrane-like locality-preserving transformations. In this subsection, we present complete classifications of transparent gapped domain walls for the two-dimensional toric code and the two-dimensional color code. Namely, we find that every transparent domain wall has a corresponding membrane operator associated with non-trivial automorphism of anyon labels. 

In $(2+1)$-dimensional TQFTs, labels of anyons, which may condense on a gapped boundary, can be characterized by maximal sets of anyons with trivial self and mutual braiding statistics~\cite{Levin12}. To characterize a gapped domain wall which connects two topologically ordered systems, we fold the system along the domain wall and view the domain wall as a gapped boundary where two systems (the left and the right) are attached together~\cite{Beigi11}. We are then able to classify automorphisms of anyon labels on the domain wall as condensation of anyons from the left and the right in the folded geometry. 

To begin, let us characterize transparent domain walls of the toric code. Anyons are denoted by $e,m$. One finds the following two transparent domain walls:
\begin{equation}
\begin{split}
&W_{0} : (e|e),\quad (m|m) \\
&W_{1} : (e|m),\quad (m|e).
\end{split}
\end{equation}
Here $W_{0}$ represents a trivial domain wall where anyon labels are not altered and $W_{1}$ corresponds to the aforementioned domain wall in the toric code which exchanges $e$ and $m$. One can view the domain wall $W_{1}$ as a gapped boundary which absorbs $e_{\ell}m_{r}$ and $m_{\ell}e_{r}$ where the subscripts $\ell$ and $r$ denote the left and the right. One then sees that $e_{\ell}m_{r}$ and $m_{\ell}e_{r}$ have trivial self and mutual braiding statistics. We note that this classification of transparent domain walls in the toric code is well known in the literature, see~\cite{Lan15} for instance.

Next, let us characterize transparent domain walls for the two-dimensional color code, which is unitarily equivalent to two copies of the toric code. The problem can be reduced to finding $4\times 4$ matrices with binary entries which satisfy certain conditions reflecting triviality of braiding statistics in the folded geometry~\cite{Lan15}. We found $72$ different types of transparent domain walls in the color code by an exhaustive search. Observe that transparent domain walls form a group since a product of two walls is also a wall. As such, domain walls can be constructed from a certain complete set of generators of domain walls. Below, we list five types of domain walls which form a complete set. 
\begin{equation}
\begin{split}
&W_{1} : (e_{A}|m_{A}),\ (m_{A}|e_{A}m_{A}),\ (e_{B}|m_{B}),\ (m_{B}|e_{B}m_{B}) \\
&W_{2} : (e_{A}|m_{A}),\ (m_{A}|e_{A}),\ (e_{B}|m_{B}),\ (m_{B}|e_{B}).
\end{split}
\end{equation}
These domain walls preserve the color labels of anyons while exchanging electric charges and magnetic fluxes. Here $W_{1}$ corresponds to $T$ operators which permute Pauli $X,Y,Z$ operators and $W_{2}$ corresponds to Hadamard operators. Here $W_{1}$ and $W_{2}$ form a subgroup which is isomorphic to the symmetric group $S_{3}$. Thus the group of domain walls is non-abelian. The following domain walls exchange color labels of anyons:
\begin{equation}
\begin{split}
&W_{3} : (e_{A}|e_{B}),\ (m_{A}|m_{B}),\ (e_{B}|e_{A}),\ (m_{B}|m_{A})\\
&W_{4} : (e_{A}|e_{B}),\ (m_{A}|m_{B}),\ (e_{B}|e_{A}e_{B}),\ (m_{B}|m_{A}m_{B}).
\end{split}
\end{equation}
One can construct the corresponding membrane operators by decoupling the color code into two copies of the toric code and exchanging them. Finally, we find the following domain wall
\begin{align}
W_{5} : (e_{A}|m_{B}),\ (m_{A}|m_{A}),\ (e_{B}|e_{B}),\ (m_{B}|e_{A})
\end{align}
which corresponds to a domain wall in a single copy of the toric code. We find that domain walls $W_{1},\ldots,W_{5}$ generate all the $72$ different transparent gapped domain walls in the color code. The main conclusion is that, for every transparent domain wall in the color code, there exists a corresponding membrane-like locality-preserving operator which preserves the ground state space and induces non-trivial automorphism of anyon labels. 

Classification of gapped domain walls is an important problem in condensed matter physics community as many of realistic physical systems have boundaries and domain walls. Classification of fault-tolerantly implementable logical gates is of relevance to the quantum information science community as they are indispensable building blocks for fault-tolerant quantum computation. Whether the correspondence between domain walls and membrane operators generically holds for $(2+1)$-dimensional TQFTs is an interesting future problem to study.

\section{Volume operator in three-dimensional color code}\label{sec:3dim_op}

\begin{figure}[htb!]
\centering
\includegraphics[width=0.75\linewidth]{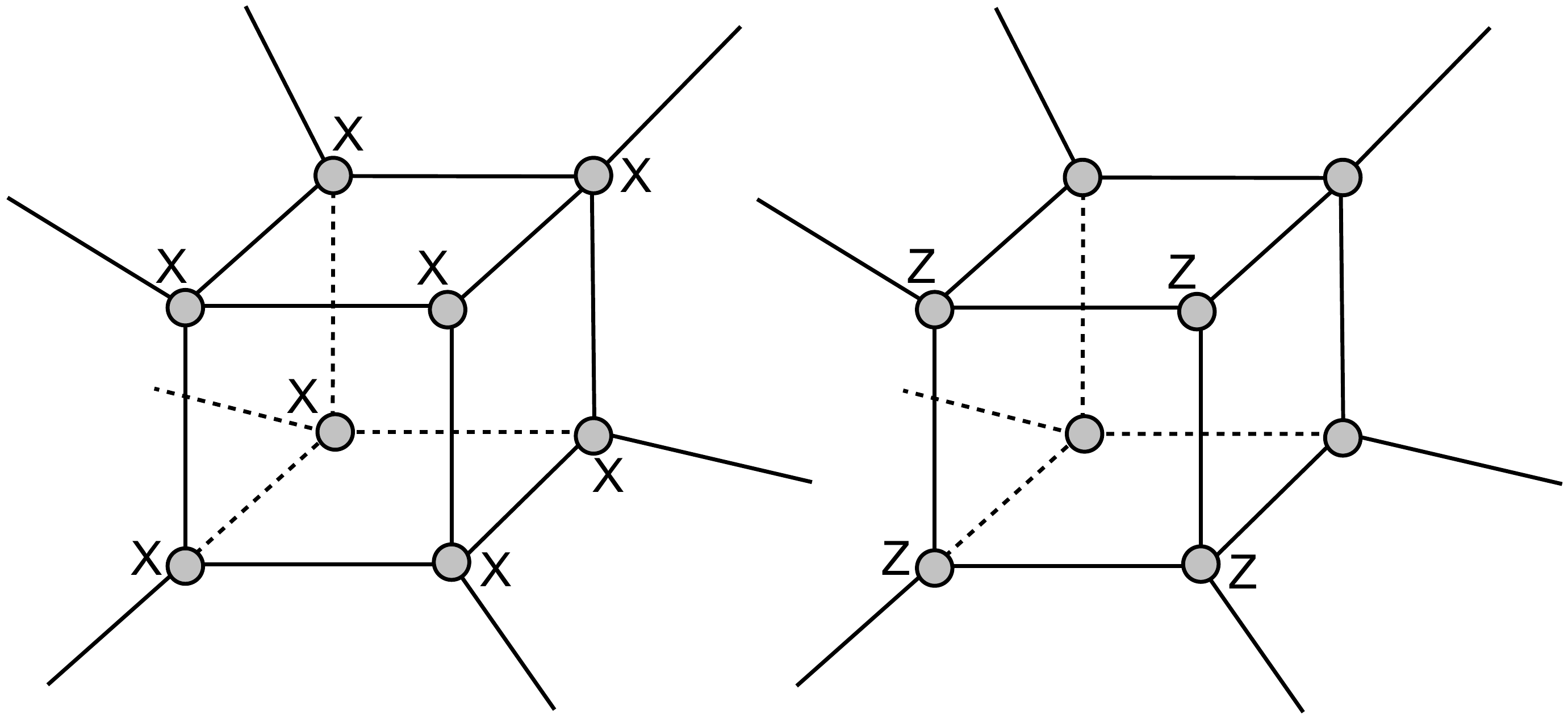}
\caption{Stabilizer operators in the three-dimensional color code. $X$-type stabilizers are associated with volumes while $Z$-types stabilizers are associated with plaquettes.  
} 
\label{fig_3dim_color}
\end{figure}

Consider the three-dimensional color code defined on a four-valent and four-colorable lattice $\Lambda$ where qubits live on vertices. Colors are denoted by $A,B,C,D$, and are associated with volumes. The Hamiltonian is given by
\begin{align}
H = - \sum_{P}S^{(Z)}_{P} -\sum_{G}S^{(X)}_{G}
\end{align}
where $P$ represents a plaquette and $G$ represents a volume (Fig.~\ref{fig_3dim_color}). Here $S^{(X)}_{G},S^{(Z)}_{P}$ commute with each other due to the four-valence and four-colorability of the lattice. For simplicity of discussion, we assume that the lattice $\Lambda$ is supported on (homomorphic to) a three-sphere so that the ground state is unique. A systematic procedure of constructing such a lattice is known~\cite{Bombin07}. 

\begin{figure}[htb!]
\centering
\includegraphics[width=0.95\linewidth]{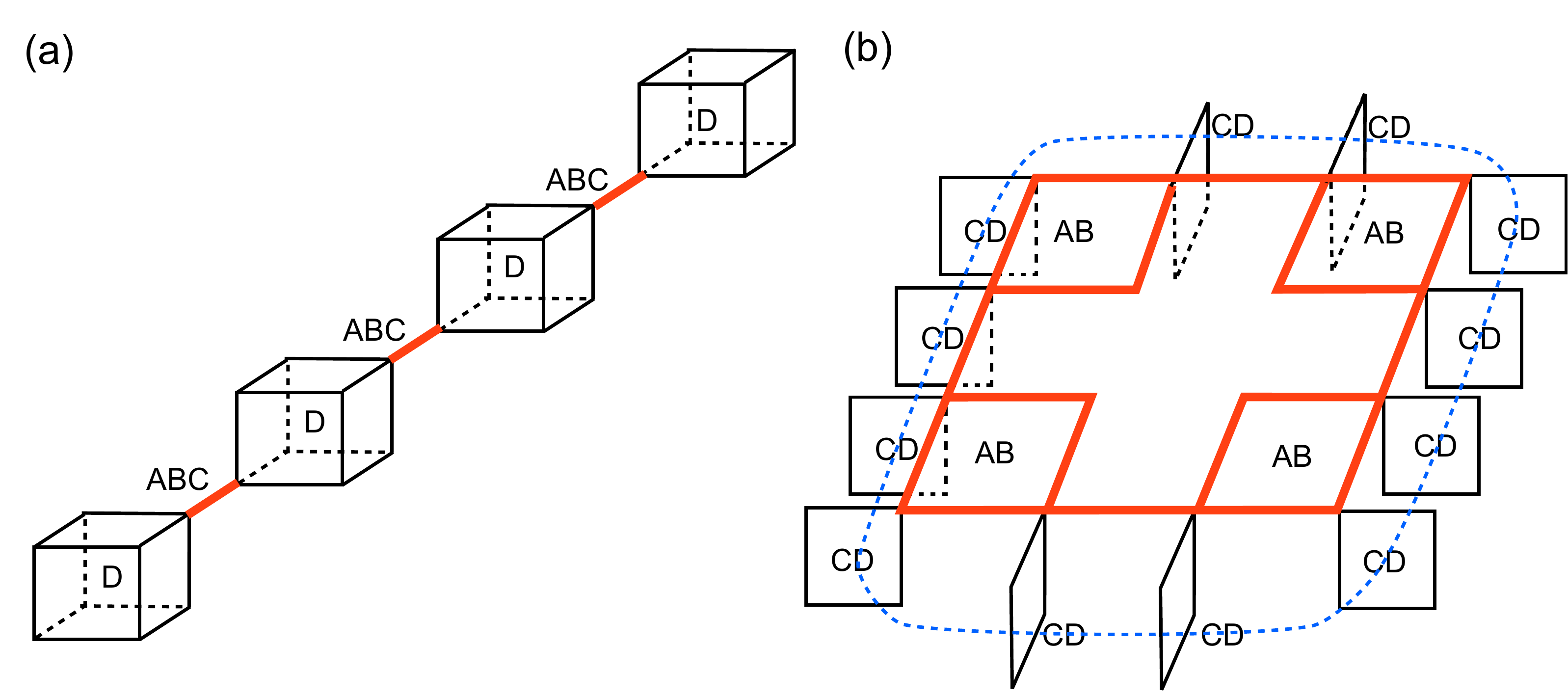}
\caption{(a) An open line $\gamma^{ABC}$, consisting of edges of color $ABC$, which defines a string-like operator $\overline{Z^{ABC}}|_{\gamma^{ABC}}$. The string connects volumes of color $D$, and creates excitations on two volumes of color $D$ sitting at the endpoints. (b) An open sheet (membrane) $\beta^{AB}$, consisting of plaquettes of color $AB$, which defines a membrane-like operator $\overline{X^{AB}}|_{\beta^{AB}}$. The membrane connects plaquettes of color $CD$, forming a loop-like flux, and creates excitations on plaquettes of color $CD$ which are on the boundary of the sheet $\beta^{AB}$. 
} 
\label{fig_3dim_excitation}
\end{figure}

Three-dimensional topologically ordered spin systems may possess both point-like and loop-like excitations which are characterized by string and two-dimensional membrane operators respectively. We begin by constructing string operators. Given a four-valent and four-colorable lattice $\Lambda$, one can assign color labels $ABC,ABD,ACD,BCD$ to its edges since, for a given edge, there always exist three volumes of different colors sharing the edge. Consider a set of edges of color $ABC$ which form a one-dimensional line $\gamma^{ABC}$ such that a string $\gamma^{ABC}$ connects volumes of color $D$ as shown in Fig.~\ref{fig_3dim_excitation}(a). We define 
\begin{align}
\overline{Z^{ABC}}|_{\gamma^{ABC}}  := \bigotimes_{j \in \gamma^{ABC}} Z_{j}.
\end{align}
If $\gamma^{ABC}$ is an open line, $\overline{Z^{ABC}}|_{\gamma^{ABC}}$ commutes with all the interaction terms except stabilizers $S^{(X)}_{G}$ on volumes of color $D$ at the endpoints of $\gamma_{ABC}$. Thus, one can characterize electric charges as follows
\begin{equation}
\begin{split}
&\overline{Z^{ABC}}|_{\gamma^{ABC}} \leadsto e_{D},\quad \overline{Z^{ABD}}|_{\gamma^{ABD}} \leadsto e_{C}\\
&\overline{Z^{ACD}}|_{\gamma^{ACD}} \leadsto e_{B},\quad \overline{Z^{BCD}}|_{\gamma^{BCD}} \leadsto e_{A}.
\end{split}
\end{equation}
To construct membrane operators, we assign color labels $AB,AC,AD,BC,BD,CD$ to plaquettes of the lattice $\Lambda$. Consider a set of plaquettes of color $AB$ which form a two-dimensional sheet (membrane) $\beta^{AB}$ as shown in Fig.~\ref{fig_3dim_excitation}(b). We define 
\begin{align}
\overline{X^{AB}}|_{\beta^{AB}}  := \bigotimes_{j \in \beta^{AB}} X_{j}.
\end{align}
If $\beta^{AB}$ is an open membrane with boundaries, the operator creates excitations associated with stabilizers $S^{(Z)}_{P}$ on plaquettes of color $CD$ on boundaries of $\beta_{AB}$. Thus, one can characterize loop-like magnetic fluxes as follows
\begin{equation}
\begin{split}
&\overline{X^{AB}}|_{\beta^{AB}} \leadsto m_{CD},\quad \overline{X^{AC}}|_{\beta^{AC}} \leadsto m_{BD} \\
&\overline{X^{AD}}|_{\beta^{AD}} \leadsto m_{BC},\quad \overline{X^{BC}}|_{\beta^{BC}} \leadsto m_{AD}\\
&\overline{X^{BD}}|_{\beta^{BD}} \leadsto m_{AC},\quad \overline{X^{CD}}|_{\beta^{CD}} \leadsto m_{AB}.
\end{split}
\end{equation}
These excitations with different color labels are not independent from each other since the following fusion channels exist
\begin{equation}
\begin{split}
&e_{A}\times e_{B}\times e_{C}\times e_{D}=1,\quad m_{AB}\times m_{CD}=1\\ 
&m_{AC}\times m_{BD}=1,\quad m_{AD}\times m_{BC}=1.
\end{split}
\end{equation}
It is convenient to construct an isomorphism between anyons of the color code and those of the three-dimensional toric code. Let $e_{1},m_{1}, e_{2},m_{2}, e_{3},m_{3}$ be anyons in three decoupled copies of the toric code. Then one has
\begin{equation}
\begin{split}
&e_{A} \leftrightarrow e_{1}, \quad e_{B} \leftrightarrow e_{2},\quad e_{C} \leftrightarrow e_{3}\\
&m_{AB} \leftrightarrow m_{3}, \quad m_{AC} \leftrightarrow m_{2}, \quad m_{BC} \leftrightarrow m_{1}.\label{eq:3dim_isomorphism}
\end{split}
\end{equation}
It is known that, on a closed manifold, the three-dimensional color code is equivalent to three decoupled copies of the toric code under a local unitary transformation~\cite{Beni11, Kubica15b}. 

Note that the three-dimensional color code has volume and membrane phase operators. Recall that the lattice $\Lambda$ is bipartite and qubits can be split into two sets $\mathcal{T}$ and $\mathcal{T}^{c}$. For a phase operator $R_{3}=\mbox{diag}(1,e^{i\pi/4})$, let us define the following transversal volume (three-dimensional) phase operator
\begin{align}
\overline{R_{3}} := \bigotimes_{j \in \mathcal{T}} R_{3}|_{j} \bigotimes_{j \in \mathcal{T}^{c}} (R_{3}|_{j})^{-1}.
\end{align}
It is known that $\overline{R_{3}}$ preserves the ground state space of the three-dimensional color code. Namely, if there is only a single ground state, one has
\begin{align}
\overline{R_{3}}|\psi_{gs}\rangle \propto |\psi_{gs}\rangle.
\end{align}
This is a rather non-trivial statement, and readers are referred to~\cite{Kubica15} for a proof. 

We then define membrane-like $R_{2}$ phase operators. Let us consider a set of plaquettes of color $AB$ which form a two-dimensional sheet $\beta^{AB}$. Recall that the sheet $\beta^{AB}$ is bipartite and qubits can be split into two sets $\beta^{AB}\cap \mathcal{T}$ and $\beta^{AB}\cap \mathcal{T}^{c}$. We define 
\begin{align}
\overline{{R_{2}}^{AB}}|_{\beta^{AB}}  := \bigotimes_{j \in \beta^{AB}\cap \mathcal{T} } R_{2}|_{j}\bigotimes_{j \in \beta^{AB}\cap \mathcal{T}^{c} } (R_{2}|_{j})^{-1}.
\end{align}
If $\beta^{AB}$ is an open membrane with boundaries, the operator commutes with all the stabilizers except $X$-type stabilizers $S_{G}^{(X)}$ of color $C$ and $D$ on boundaries. So, $\overline{{R_{2}}^{AB}}|_{\beta^{AB}}$ creates a loop-like excitation involving both $e_{C}$ and $e_{D}$. This membrane-like phase operator $\overline{{R_{2}}^{AB}}|_{\beta^{AB}}$ is closely related to the other membrane operator $\overline{X^{AB}}|_{\beta^{AB}}$ by the following relation:
\begin{align}
K(\overline{R_{3}}, \overline{X^{AB}}|_{\beta^{AB}}) \propto \overline{{R_{2}}^{AB}}|_{\beta^{AB}} \label{eq:commutator}
\end{align}
where $K(U,V):=UVU^{\dagger}V^{\dagger}$ is the so-called group commutator. One can see that the excited wavefunction, resulting from  $\overline{{R_{2}}^{AB}}|_{\beta^{AB}}$, corresponds to a one-dimensional SPT wavefunction with $\mathbb{Z}_{2}\otimes \mathbb{Z}_{2}$ symmetry in the excitation basis where $\mathbb{Z}_{2}\otimes \mathbb{Z}_{2}$ symmetries are associated with parity conservations of $e_{C}$ and $e_{D}$. We shall denote such a loop-like SPT excitation by $s_{CD}$. Loop-like SPT excitations, created by $R_{2}$ membrane operators, are characterized by 
\begin{equation}
\begin{split}
&\overline{{R_{2}}^{AB}}|_{\beta^{AB}} \leadsto s_{CD},\quad \overline{{R_{2}}^{AC}}|_{\beta^{AC}} \leadsto s_{BD} \\
&\overline{{R_{2}}^{AD}}|_{\beta^{AD}} \leadsto s_{BC},\quad \overline{{R_{2}}^{BC}}|_{\beta^{BC}} \leadsto s_{AD}\\
&\overline{{R_{2}}^{BD}}|_{\beta^{BD}} \leadsto s_{AC},\quad \overline{{R_{2}}^{CD}}|_{\beta^{CD}} \leadsto s_{AB}
\end{split}
\end{equation}
with the following fusion channels
\begin{align}
s_{AB}\times s_{CD}=1,\  s_{AC}\times s_{BD}=1,\  s_{AD}\times s_{BC}=1.
\end{align}
For instance, if $s_{AB}$ and $s_{CD}$ are located next to each other, then one can eliminate them by local unitary transformations. 

\section{SPT excitations with $\mathbb{Z}_{2}\otimes \mathbb{Z}_{2} \otimes \mathbb{Z}_{2}$ symmetry}\label{sec:3dim_ham}

In this section, we study membrane-like excitations created by applying the phase operator $\overline{R_{3}}$ in some connected region of qubits. 
Namely, we show that membrane excitations are characterized by a wavefunction of a two-dimensional bosonic SPT phase with $\mathbb{Z}_{2}\otimes \mathbb{Z}_{2}\otimes \mathbb{Z}_{2}$ symmetry. We also find a new simple fixed-point Hamiltonian for SPT phases with $\mathbb{Z}_{2}\otimes \mathbb{Z}_{2}\otimes \mathbb{Z}_{2}$ symmetry. We conjecture that the corresponding SPT phase is the so-called type-III phase which is dual to the non-abelian quantum double model based on a dihedral group $D_{4}$. 

\subsection{Membrane excitations}

Consider a set of volumes of color $D$ which forms a connected contractible region of qubits $V$ with a single boundary. We consider a restriction of the three-dimensional phase operator $\overline{R_{3}}$ on $V$, denoted by $R_{3}|_{V}$:
\begin{align}
R_{3}|_{V} := \bigotimes_{j \in V\cap \mathcal{T}} R_{3}|_{j} \bigotimes_{j \in V\cap \mathcal{T}^{c}} (R_{3}|_{j})^{-1}.
\end{align}
Consider an excited wavefunction $|\psi_{V}\rangle:=R_{3}|_{V}\cdot |\psi_{gs}\rangle$. Since the region $V$ is constructed from a set of volumes of color $D$, $|\psi_{V}\rangle$ involves excitations only on volumes of color $A,B,C$. Since the phase operator $\overline{R_{3}}$ is diagonal in the computational basis, it creates excitations which are associated with $X$-type stabilizers on volumes. We hope to represent $|\psi_{V}\rangle$ in the excitation basis in a way similar to the two-dimensional case. Not all the basis states are physically allowed since
\begin{align}
\prod_{P \in \mathcal{A}} S^{(X)}_{P}=\prod_{P\in \mathcal{B}} S^{(X)}_{P}=\prod_{P\in \mathcal{C}} S^{(X)}_{P} = \prod_{P \in \mathcal{D}} S^{(X)}_{P}
\end{align}
where $\mathcal{A}, \mathcal{B}, \mathcal{C}, \mathcal{D}$ represent sets of volumes of color $A,B,C,D$ respectively. Let $N_{\mathcal{A}}$, $N_{\mathcal{B}}$, $N_{\mathcal{C}}$, $N_{\mathcal{D}}$ be the total number of excitations on volumes of color $A,B,C,D$. Since $N_{\mathcal{D}}=0$, one has
\begin{align}
N_{\mathcal{A}} = N_{\mathcal{B}}= N_{\mathcal{C}}=0 \qquad \mbox{(mod $2$)} \label{eq:3d_parity}
\end{align} 
which leads to $\mathbb{Z}_{2}\otimes \mathbb{Z}_{2}\otimes \mathbb{Z}_{2}$ symmetry of $|\psi_{V}\rangle$ in the excitation basis. Clearly, the wavefunction $|\psi_{V}\rangle$ has excitations only on the boundary $\partial V$ of $V$ since $\overline{R_{3}}|\psi_{gs}\rangle \propto|\psi_{gs}\rangle$. So, one can characterize the excitation wavefunction by the boundary wavefunction $|\phi^{(\textrm{ex})}_{\partial V}\rangle$ which corresponds to the presence and absence of excitations associated with volumes on the boundary $\partial V$ of $V$.

\begin{figure}[htb!]
\centering
\includegraphics[width=0.95\linewidth]{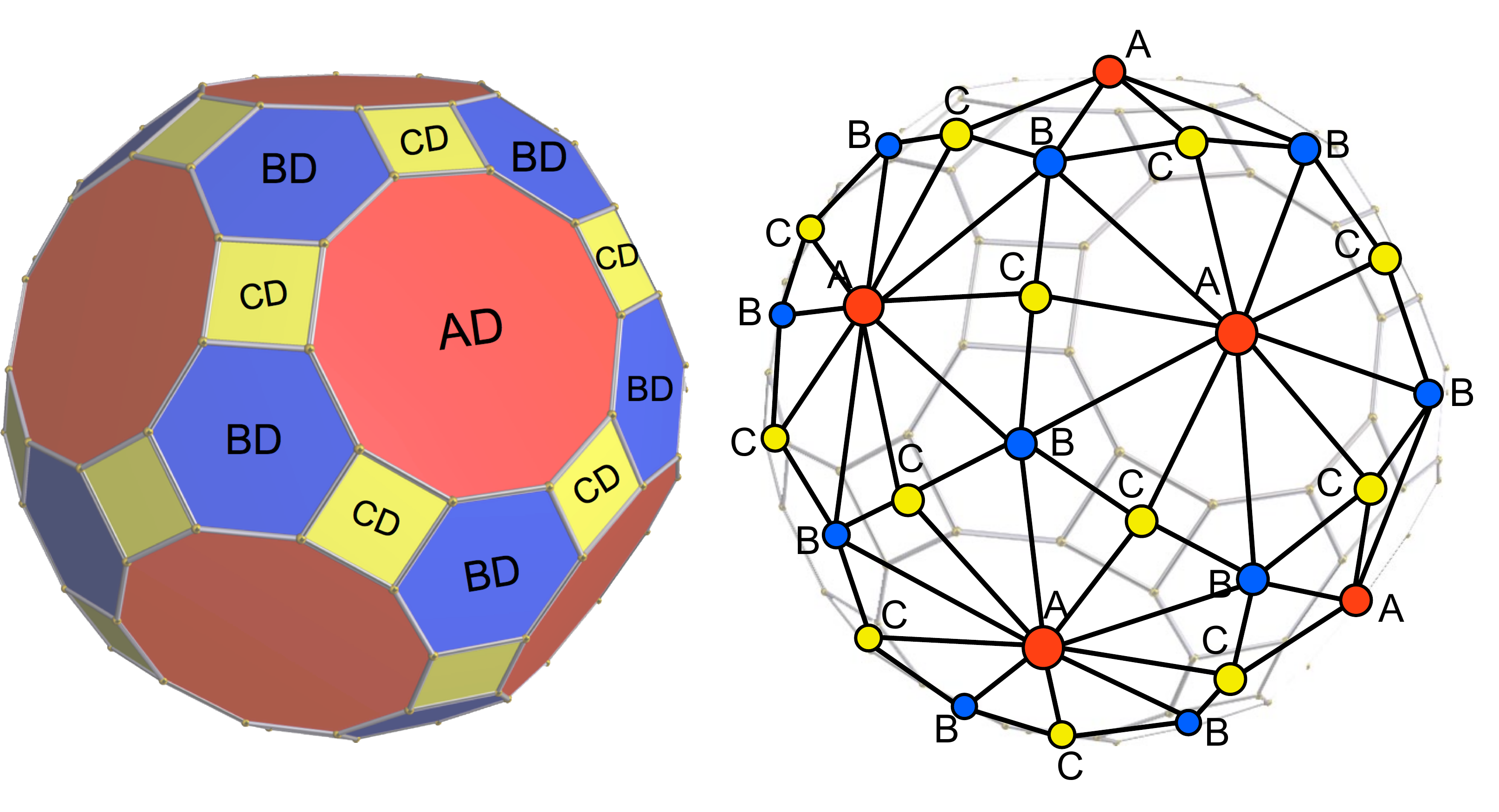}
\caption{The boundary of a volume of color $D$, viewed as a three-colorable lattice consisting of plaquettes of color $AD,BD,CD$. The boundary excitation wavefunction $|\phi^{(\textrm{ex})}_{\partial V}\rangle$ is supported on qubits associated with three-colorable vertices on the dual lattice $\partial V^{\star}$. The figures were created using Robert WebbÕs Stella software (http://www.software3d.com/Stella.php).
} 
\label{fig_surface}
\end{figure}

\begin{figure}[htb!]
\centering
\includegraphics[width=0.85\linewidth]{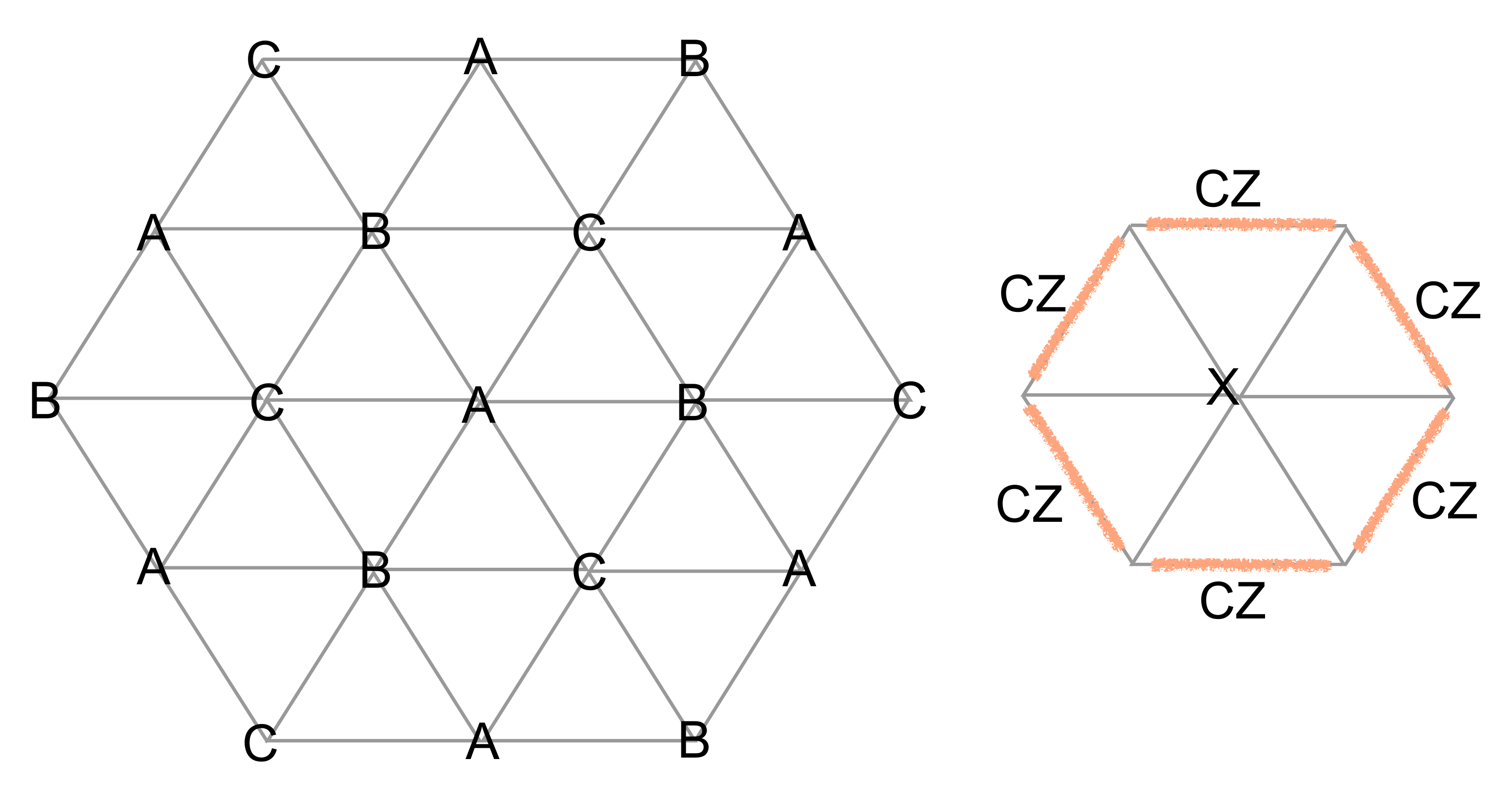}
\caption{The Hamiltonian for an SPT phase with $\mathbb{Z}_{2}\otimes \mathbb{Z}_{2}\otimes \mathbb{Z}_{2}$ symmetry, supported on a triangular lattice with vertices of three different colors. 
} 
\label{fig_3dim_model}
\end{figure}

Let us find the boundary wavefunction $|\phi^{(\textrm{ex})}_{\partial V}\rangle$. 
A key observation is that the boundary $\partial V$ can be viewed as a three-valent and three-colorable lattice (see Fig.~\ref{fig_surface}). 
Namely, since $V$ is a set of volumes of $D$, the boundary $\partial V$ consists of plaquettes of color $AD,BD,CD$. 
Consider volumes of color $A,B,C$ which are located on the boundary $\partial V$. 
Such volumes on the boundary, which form the basis states of $|\phi^{(\textrm{ex})}_{\partial V}\rangle$, can be associated with plaquettes of color $AD,BD,CD$ on the boundary $\partial V$. 
As such, the boundary wavefunction $|\phi^{(\textrm{ex})}_{\partial V}\rangle$ can be interpreted as a two-dimensional wavefunction which involve qubits supported on plaquettes of a three-valent and three-colorable lattice $\partial V$. Here, it is convenient to introduce a dual lattice picture (Fig.~\ref{fig_surface}). Let us construct a dual boundary lattice $\partial V^{\star}$ by viewing the centers of plaquettes in $\partial V$ as vertices. Then, in a dual lattice $\partial V^{\star}$, qubits live on vertices and vertices are colored in $A,B,C$. In summary, the boundary wavefunction $|\phi^{(\textrm{ex})}_{\partial V}\rangle$ can be viewed as a two-dimensional system of qubits supported on a triangular lattice without boundaries whose vertices are colored in $A,B,C$ (Fig.~\ref{fig_3dim_model}). 

One can find that the boundary wavefunction $|\phi^{(\textrm{ex})}_{\partial V}\rangle$ is given by 
\begin{align}
|\phi^{(\textrm{ex})}_{\partial V}\rangle = \prod_{\langle i,j,k\rangle} \exp\left( \pm \frac{i \pi}{8} X_{i}X_{j}X_{k} \right) |0\rangle^{\otimes n} \label{eq:3d_state}
\end{align}
where $n$ represents the number of vertices and $\langle i,j,k\rangle$ represents a triangle in the dual lattice $\partial V^{\star}$. Here ``$\pm$'' in the product corresponds to $R_{3}$ and $(R_{3})^{-1}$ in the phase operator $\overline{R_{3}}$ respectively. One can construct a Hamiltonian which has this wavefunction $|\phi^{(\textrm{ex})}_{\partial V}\rangle$ as a unique ground state: 
\begin{align}
H = - \sum_{j} \Big( Z_{j} \prod_{\langle j qq' \rangle} \exp(i\pi t_{q}t_{q'})\Big)\label{eq:3d_hamiltonian}
\end{align}
where the product runs over triangles $\langle jqq' \rangle$ containing the vertex $j$, and
\begin{align}
t_{q}:=\frac{X_{q}+1}{2}.
\end{align}
Here $X_{q}$ is a Pauli-$X$ operator acting on a qubit at a vertex $q$. The boundary wavefunction is a unique gapped ground state of this Hamiltonian (assuming $\partial V$ is a closed manifold):
\begin{align}
Q_{j}|\phi^{(\textrm{ex})}_{\partial V}\rangle  = + |\phi^{(\textrm{ex})}_{\partial V}\rangle \quad \forall j
\end{align} 
and has $\mathbb{Z}_{2}\otimes \mathbb{Z}_{2}\otimes \mathbb{Z}_{2}$ symmetry emerging from parity conservation in Eq.~(\ref{eq:3d_parity}).

\subsection{SPT phase with $\mathbb{Z}_{2}\otimes \mathbb{Z}_{2}\otimes \mathbb{Z}_{2}$ symmetry}

In this subsection, we further establish the connection between the boundary wavefunction and SPT phases with $\mathbb{Z}_{2}\otimes \mathbb{Z}_{2}\otimes \mathbb{Z}_{2}$ symmetry. To be consistent with the notation used in the literature~\cite{Chen13, Levin12}, we shall apply transversal Hadamard operators to the wavefunction and the Hamiltonian in Eq.~(\ref{eq:3d_state}) and Eq.~(\ref{eq:3d_hamiltonian}). After exchanging $X$ and $Z$, the wavefunction is given by (Fig.~\ref{fig_3dim_model})
\begin{align}
|\psi_{\textrm{SPT}} \rangle := \prod_{\langle i,j,k\rangle} \exp\left( \pm \frac{i \pi}{8} Z_{i}Z_{j}Z_{k} \right) |+\rangle^{\otimes n}
\end{align} \label{eq:state}
and the Hamiltonian is given by 
\begin{align}
H_{\textrm{SPT}} := - \sum_{j} \Big(X_{j} \prod_{\langle j qq' \rangle} \mbox{CZ}_{qq'}\Big) \label{eq:hamiltonian}
\end{align}
where the so-called control-$Z$ operator $\mbox{CZ}_{qq'}$ is given by
\begin{align}
\mbox{CZ}_{qq'}:=\exp(i\pi s_{q}s_{q'}) \quad s_{q}:=\frac{Z_{q}+1}{2}.
\end{align}
(Here $\mbox{CZ}_{qq'}$ applies a Pauli $Z$ operator on $q'$ provided the qubit at $q$ is in a state $|1\rangle_{q}$). The wavefunction satisfies the stabilizer conditions:
\begin{align}
h_{j}|\psi_{\textrm{SPT}} \rangle=+ |\psi_{\textrm{SPT}} \rangle \quad \forall j.
\end{align}
Let us define symmetry operators as follows:
\begin{align}
\mathcal{S}_{A} := \bigotimes_{j} X^{(A)}_{j},\ \mathcal{S}_{B} := \bigotimes_{j} X^{(B)}_{j},\ \mathcal{S}_{C} := \bigotimes_{j} X^{(C)}_{j}
\end{align}
which are associated with parity conservation. One can verify that the wavefunction is symmetric:
\begin{align}
\mathcal{S}_{A}|\psi_{\textrm{SPT}} \rangle = \mathcal{S}_{B}|\psi_{\textrm{SPT}} \rangle  = \mathcal{S}_{C}|\psi_{\textrm{SPT}} \rangle  = |\psi_{\textrm{SPT}} \rangle.
\end{align} 
To see this, notice that the following relations hold
\begin{align}
\prod_{j\in \mathcal{A}}Q_{j}=\mathcal{S}_{A},\quad \prod_{j\in \mathcal{B}}Q_{j}=\mathcal{S}_{B}, \quad \prod_{j\in \mathcal{C}}Q_{j}=\mathcal{S}_{C}
\end{align}
which can be shown by recalling the fact that $(\mbox{CZ}_{qq'})^2=I$. Thus, the boundary wavefunction has $\mathbb{Z}_{2}\otimes \mathbb{Z}_{2}\otimes \mathbb{Z}_{2}$ symmetry.

Readers who are familiar with the literature on bosonic SPT phases may notice that the aforementioned Hamiltonian is identical to the fixed-point Hamiltonian of two-dimensional SPT phases with $\mathbb{Z}_{2}$ symmetry proposed in Ref.~\cite{Levin12, Chen13}, \emph{instead of} $\mathbb{Z}_{2}\otimes \mathbb{Z}_{2}\otimes \mathbb{Z}_{2}$ symmetry. The crucial observation is that this Hamiltonian can be also viewed as a fixed-point Hamiltonian of two-dimensional SPT phases with $\mathbb{Z}_{2}\otimes \mathbb{Z}_{2}\otimes \mathbb{Z}_{2}$ symmetry if an appropriate set of symmetry operators is considered. In Ref.~\cite{Levin12}, the Hamiltonian in Eq.~(\ref{eq:hamiltonian}), supported on a triangular lattice, was treated as a $\mathbb{Z}_{2}$ symmetric model with respect to $\mathcal{S}_{A}\mathcal{S}_{B}\mathcal{S}_{C}$. Yet, if one views the same Hamiltonian with $\mathbb{Z}_{2}\otimes \mathbb{Z}_{2}\otimes \mathbb{Z}_{2}$ symmetry with respect to $\mathcal{S}_{A}\otimes \mathcal{S}_{B}\otimes \mathcal{S}_{C}$, the system can be viewed as an example of SPT phases with $\mathbb{Z}_{2}\otimes \mathbb{Z}_{2}\otimes \mathbb{Z}_{2}$ symmetry. This change of symmetry operators has a highly non-trivial consequence since the resulting $\mathbb{Z}_{2}\otimes \mathbb{Z}_{2}\otimes \mathbb{Z}_{2}$ symmetric wavefunction is different from three decoupled copies of $\mathbb{Z}_{2}$ symmetric wavefunctions. 

We then argue that the wavefunction $|\psi_{\textrm{SPT}} \rangle$ belongs to a non-trivial SPT phase with $\mathbb{Z}_{2}\otimes \mathbb{Z}_{2}\otimes \mathbb{Z}_{2}$ symmetry. We present two different arguments to reach this conclusion. First, one can consider a thought experiment of creating a magnetic flux and making it cross this excitation as in two-dimensional case. As discussed further in the next section, a magnetic flux becomes a composite of magnetic flux and a loop-like SPT excitation upon crossing the membrane excitation. This implies that membrane excitations associated with $R_{3}$ operators cannot be locally created by applying local unitary transformation along the boundary. The second argument replies on the fact that  $|\psi_{\textrm{SPT}} \rangle$ belongs to a non-trivial SPT with $\mathbb{Z}_{2}$ symmetry, which has been already established in Ref.~\cite{Levin12} by gauging the on-site symmetries. This implies that $|\psi_{\textrm{SPT}} \rangle$ must be non-trivial with respect to $\mathbb{Z}_{2}\otimes \mathbb{Z}_{2}\otimes \mathbb{Z}_{2}$ too. It remains to determine which class of SPT phases $|\psi_{\textrm{SPT}} \rangle$ corresponds to. 

It is known that there are $128$ different SPT phases with $\mathbb{Z}_{2}\times \mathbb{Z}_{2}\times \mathbb{Z}_{2}$ symmetry including the trivial one~\cite{Chen13, Propitius95}. These $128$ different phases can be fully characterized by three different types of SPT phases called type-I, type-II and type-III. 
Their corresponding cocycles are given by
\begin{equation}
\begin{split}
\omega_{I}(A,B,C)&= \exp\left(\frac{ i\pi }{2} a^{(i)}(b^{(i)} + c^{(i)} - [b^{(i)} + c^{(i)}])  \right) \\
\omega_{II}(A,B,C)&= \exp\left(\frac{i\pi }{2} a^{(i)}(b^{(i)} + c^{(i)} - [b^{(i)} + c^{(i)}])  \right) \\
\omega_{III}(A,B,C)&= \exp\left( i\pi  a^{(i)}b^{(j)} c^{(\ell)} \right)
\end{split}
\end{equation}
where $1\leq i<j<\ell\leq 3$. Note $A=(a^{(1)},a^{(2)},a^{(3)})$, $B=(b^{(1)},b^{(2)},b^{(3)})$, $C=(c^{(1)},c^{(2)},c^{(3)})$, and $a^{(j)},b^{(j)},c^{(j)}=0,1$. Here brackets represent modulo $2$ calculus. 
By gauging on-site symmetries, these three types of SPT phases are mapped to the twisted quantum double models which are spin systems with intrinsic topological order (without symmetries).  The topological model associated with type-III is known to be dual to the (untwisted) quantum double model based on the dihedral group $D_{4}$ with non-abelian topological order. We conjecture that the above Hamiltonian with $\mathcal{S}_{A}\otimes \mathcal{S}_{B}\otimes \mathcal{S}_{C}$ symmetry corresponds to the type-III class since the wavefunction mixes three different modes $A,B,C$ and possesses three-party entanglement among $A,B,C$. Another argument supporting this conjecture relies on the consideration of boundaries. As discussed in a forthcoming paper~\cite{Beni15}, in the $d$-dimensional quantum double model based on a finite group $G$, gapped boundaries can be constructed by considering $d$-cocycles with respect to $H\subseteq G$. For three decoupled copies of the three-dimensional toric code (\emph{i.e.} the three-dimensional quantum double model with $\mathbb{Z}_{2}\otimes \mathbb{Z}_{2}\otimes \mathbb{Z}_{2}$ symmetry), one can construct a transparent domain wall by using $3$-cocyle of $\mathbb{Z}_{2}\otimes \mathbb{Z}_{2}\otimes \mathbb{Z}_{2} \subseteq (\mathbb{Z}_{2}\otimes \mathbb{Z}_{2}\otimes \mathbb{Z}_{2})^{\otimes 2}$ which corresponds to the type-III class. One can verify that such a transparent domain wall transforms excitations as described in Eq.~(\ref{eq:wall1}) and Eq.~(\ref{eq:wall2}). It is also worth noticing that operators $X$ and $R_{2}$ form a projective representation of $D_{4}$. We note that one may also identify the corresponding SPT phase by considering symmetry twist or flux insertion as proposed in Ref.~\cite{Zaletel14, Hung14, J_Wang14, Wang15}. 

To conclude this section, we briefly mention excitations in the higher-dimensional color code. There exists a $d$-dimensional generalization of the topological color code with point-like electric charge and codimension-$1$ magnetic flux. The code admits the transversal $R_{k}$ phase operators (for $1<k\leq d$) which are supported on $k$-dimensional regions and preserve the ground state space. One can then study codimension-$1$ excitations created by $d$-dimensional $R_{d}$-type operators, which lead to a $(d-1)$-dimensional boundary wavefunction with $(\mathbb{Z}_{2})^{\otimes d}$ symmetry. The emerging model is supported on a $(d-1)$-dimensional lattice where qubits are associated with vertices which are $d$-colorable. The Hamiltonian for this wavefunction involves a Pauli $X$ operator and generalized $(d-1)$-qubit control-$Z$ operators~\cite{Kubica15b}, and the colorability of the lattice is essential to ensure that the system has $(\mathbb{Z}_{2})^{\otimes d}$ symmetry.

\section{Two-dimensional boundary and three-loop braiding}\label{sec:3dim_wall}

\subsection{The $R_{3}$ boundary}

In $(2+1)$-dimensional TQFTs, transparent gapped domain walls are classified by automorphisms of anyonic excitations. The most important distinction between two-dimensional and three-dimensional systems is that transparent domain walls in $(3+1)$-dimensional TQFTs \emph{cannot} be classified by transpositions of eigenstate excitations such as electric charges and magnetic fluxes. In this subsection, we demonstrate that in the three-dimensional color code a transparent domain wall may transform a loop-like magnetic flux into a composite of a magnetic flux and a loop-like SPT excitation. 

\begin{figure}[htb!]
\centering
\includegraphics[width=0.85\linewidth]{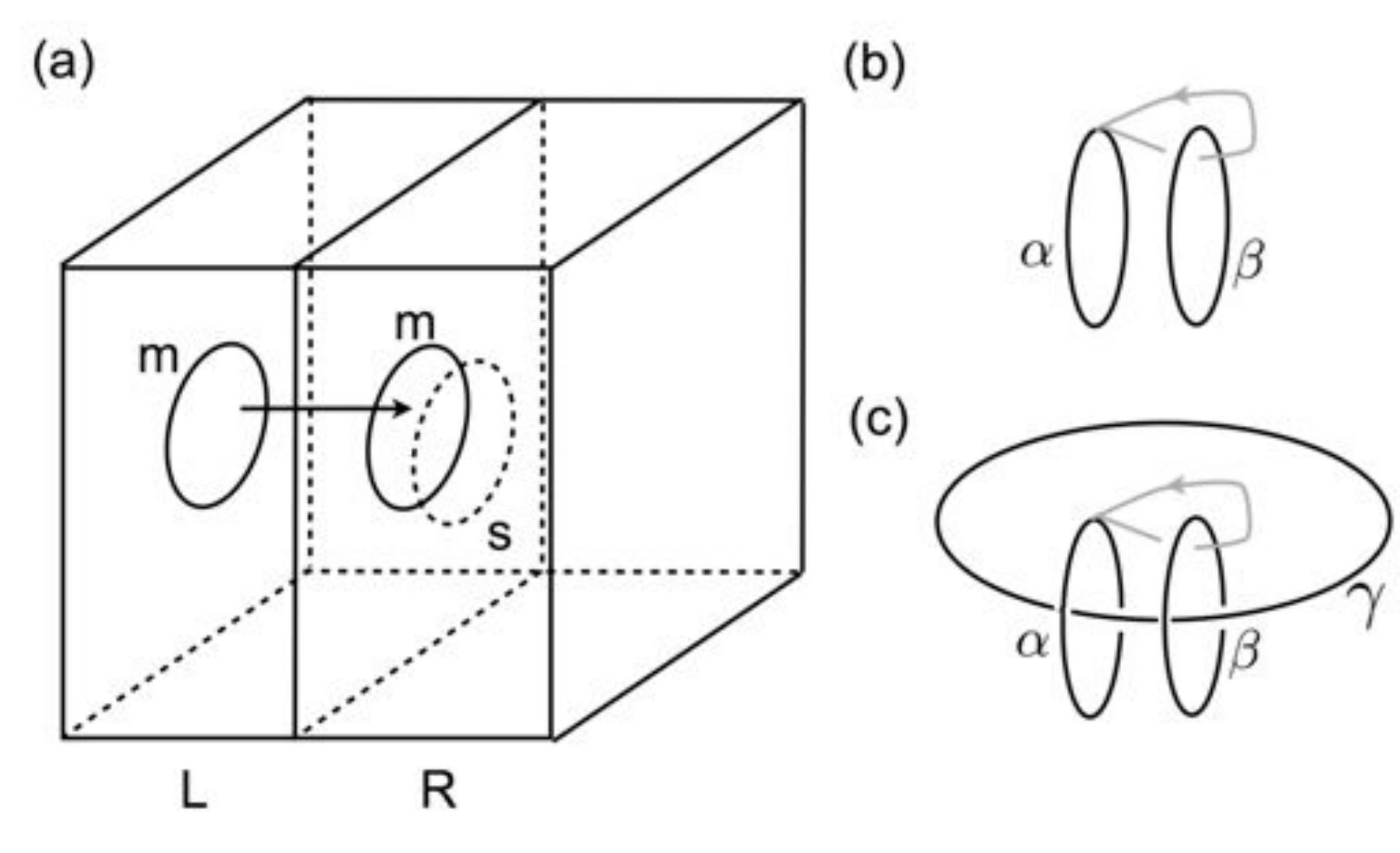}
\caption{(a) A transparent domain wall created by $R_{3}$ phase operators applied to the right part $R$ of the system.
(b) Two-loop braiding process. (c) Three-loop braiding process. 
} 
\label{fig_3dim_wall}
\end{figure}

As in the two-dimensional case, we split the entire system into two parts, left $L$ and right $R$. We then apply $R_{3}$ operators only on qubits in $R$. This creates a gapped domain wall on the boundary between $L$ and $R$ (Fig.~\ref{fig_3dim_wall}). Since the $R_{3}$ phase operator commutes with Pauli-$Z$ operators, electric charges cross the domain wall without being affected:
\begin{align}
(e_{A}|e_{A}),\ (e_{B}|e_{B}),\ (e_{C}|e_{C}).\label{eq:wall1}
\end{align}
Yet, magnetic fluxes get transformed in an interesting way. Recall that $R_{3}X  R_{3}^{\dagger}\propto X R_{2}$. Then, one finds that a magnetic flux becomes a composite of a magnetic flux \emph{and} a one-dimensional SPT excitation with $\mathbb{Z}_{2}\otimes \mathbb{Z}_{2}$ symmetry. Namely, one has
\begin{equation}
\begin{split}
&(m_{AB}|m_{AB}s_{AB}),\ (m_{BC}|m_{BC}s_{BC}),\ (m_{CA}|m_{CA}s_{CA})\\ 
&(s_{AB}|s_{AB}),\ (s_{BC}|s_{BC}),\ (s_{CA}|s_{CA}). \label{eq:wall2}
\end{split}
\end{equation}
Thus, in order to characterize gapped domain walls in the three-dimensional color code, SPT excitations need to be considered. 

\subsection{Braiding statistics between two excitations}

In $(2+1)$-dimensional TQFTs, anyons which may condense on the boundary can be classified by considering maximal sets of mutually bosonic anyons~\cite{Levin12,Barkeshli13,Levin13}. Their argument can be generalized to higher-dimensional systems too, leading to a conclusion that excitations which may condense on boundaries must have mutually trivial braiding statistics. Thus, in order to classify boundaries and domain walls in $(3+1)$-dimensional TQFTs, braiding statistics among particles and loops need to be studied. In this subsection and the next, we study braiding processes among electric charges, magnetic fluxes and SPT excitations in the three-dimensional color code. This subsection is devoted to braiding statistics of two excitations (Fig.~\ref{fig_3dim_wall}(b)) while the next subsection is devoted to three-loop braiding statistics (Fig.~\ref{fig_3dim_wall}(c)).

In three-dimensional systems, there are three possible braiding processes which involve two excitations: (i) particle-particle, (ii) particle-loop and (iii) loop-loop. By braiding, we mean a process of creating excitations, wind one around the other, and annihilating them separately. Here we are interested in the resulting $U(1)$ statistical phase. First, consider braiding between electric charges. Since electric charges $e_{A},e_{B},e_{C}$ are mutually bosonic, one has
\begin{align}
e^{i\theta(e_{A},e_{B})}=1
\end{align}
where $\theta(e_{A},e_{B})$ represents the corresponding $U(1)$ statistical phases. The above equation also holds under permutations of color labels. 

Second, consider braiding of a particle and a loop-like excitation. Braiding statistics is given by
\begin{align}
e^{i\theta(e_{A},m_{BC})}=-1, \  e^{i\theta(e_{A},m_{AB})}=e^{i\theta(e_{A},m_{CA})}=1.
\end{align}
So, electric charge $e_{A}$ and a magnetic flux $m_{BC}$ have non-trivial braiding statistics with additional $-1$ phase factor. This is because propagations of $e_{A}$ are characterized by a string-like Pauli-$Z$ operator $\overline{Z^{BCD}}_{\gamma^{BCD}}$ while propagations of $m_{BC}$ are characterized by a membrane-like Pauli $X$ operator $\overline{X^{AD}}_{\beta^{AD}}$. Namely, a string $\gamma^{BCD}$ and a membrane $\beta^{AD}$ overlap with each other odd times. The braiding of SPT excitations and an electric charges is trivial since SPT excitations are superpositions of electric charges:
\begin{align}
e^{i\theta(e_{A},s_{AB})}=e^{i\theta(e_{A},s_{BC})}=1.
\end{align}

\begin{figure}[htb!]
\centering
\includegraphics[width=0.65\linewidth]{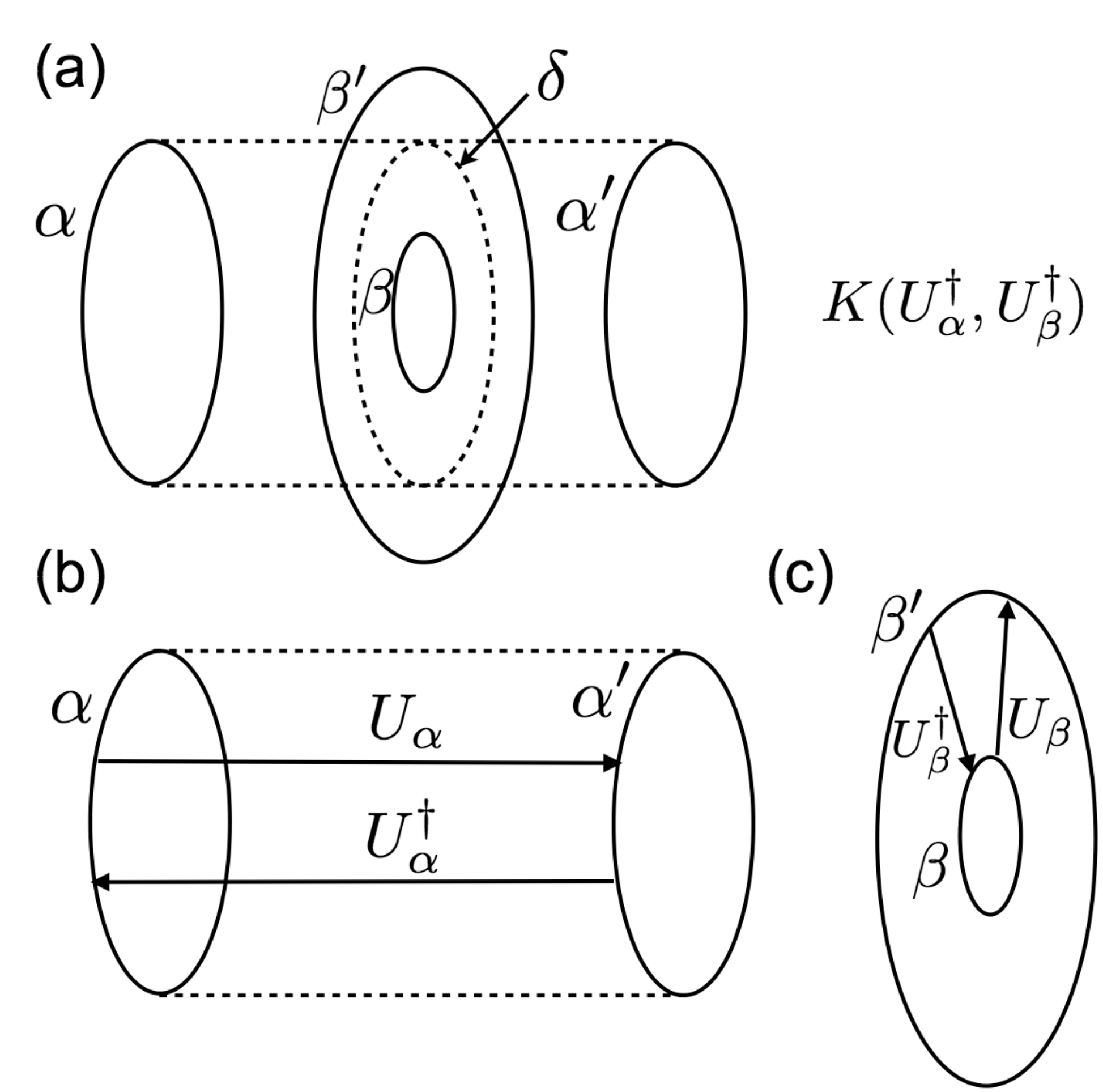}
\caption{(a) Two-loop braiding process as a group commutator $K(U_{\alpha}^{\dagger},U_{\beta}^{\dagger})$. (b) Pair of loop excitations $\alpha,\alpha'$. (c) Pair of loop excitations $\beta,\beta'$.
} 
\label{fig_two_braiding}
\end{figure}

Third, consider braiding of two loop-like excitations. In general, braiding statistics of two loop-like excitations $\alpha$ and $\beta$ can be probed by a process depicted in Fig.~\ref{fig_two_braiding}(a). Let $U_{\alpha}$ be a unitary operator corresponding to creating a pair of loops $\alpha$ and $\alpha'$, and sending $\alpha'$ in a direction perpendicular to the loop $\alpha$ as depicted in Fig.~\ref{fig_two_braiding}(b). Let $U_{\beta}$ be a unitary operator corresponding to creating a pair of loops $\beta$ and $\beta'$, and enlarge $\beta'$ as depicted in Fig.~\ref{fig_two_braiding}(c). Note that $U_{\alpha}^{\dagger}$ and $U_{\beta}^{\dagger}$ are unitary operators which reverse the processes of $U_{\alpha}$ and $U_{\beta}$. The following sequence of unitary transformations implements a braiding of two loops $\alpha$ and $\beta$ (Fig.~\ref{fig_two_braiding}(a)):
\begin{align}
U_{\alpha}^{\dagger}U_{\beta}^{\dagger}U_{\alpha}U_{\beta}|\psi_{gs}\rangle = e^{i\theta(\alpha,\beta)}|\psi_{gs}\rangle.
\end{align}
Recall that $K(U_{\alpha}^{\dagger},U_{\beta}^{\dagger})=U_{\alpha}^{\dagger}U_{\beta}^{\dagger}U_{\alpha}U_{\beta}$. As such, two-loop braiding statistics of $\alpha$ and $\beta$ can be characterized by the vacuum expectation value of the group commutator $K(U_{\alpha}^{\dagger},U_{\beta}^{\dagger})$. Note that continuous deformations of the braiding trajectories do not change the value of $e^{i\theta(\alpha,\beta)}$, and thus braiding statistics are topologically invariant.

Let us now consider a braiding of a magnetic flux $m_{AB}$ and an SPT excitation $s_{BC}$. Let $\delta$ be a loop, where the trajectories of $m_{AB}$ and $s_{BC}$ intersect (see Fig.~\ref{fig_two_braiding}(a)). Note that $U_{\alpha}$ for $\alpha=m_{AB}$ is a membrane operator with $X$ operators acting on plaquettes of color $CD$. Similarly, $U_{\beta}$ for $\alpha=s_{BC}$ is a membrane operator with $R_{2}$ operators acting on plaquettes of color $AD$. As such, the commutator $K(U_{\alpha}^{\dagger},U_{\beta}^{\dagger})$ is a loop $\delta$ of Pauli $Z$ operators acting on edges of color $ACD$. This operator acts as an identity operator in the ground state space. Thus, we conclude that 
\begin{align}
e^{i\theta(m_{AB},s_{BC})}=1.
\end{align}
Similar argument allows us to conclude that $e^{i\theta(\alpha,\beta)}=1$ for all loops $\alpha,\beta$.  

\subsection{Gapped domain wall and three-loop braiding}

Recently, Wang and Levin have proposed that full characterization of three-dimensional SPT phases requires analyzing the three-loop braiding statistics~\cite{Wang14}. In this subsection, motivated by this pioneering work, we study three-loop braiding processes among loop-like magnetic fluxes and SPT excitations and show that they exhibit non-trivial three-loop braiding statistics. 

\begin{figure}[htb!]
\centering
\includegraphics[width=0.65\linewidth]{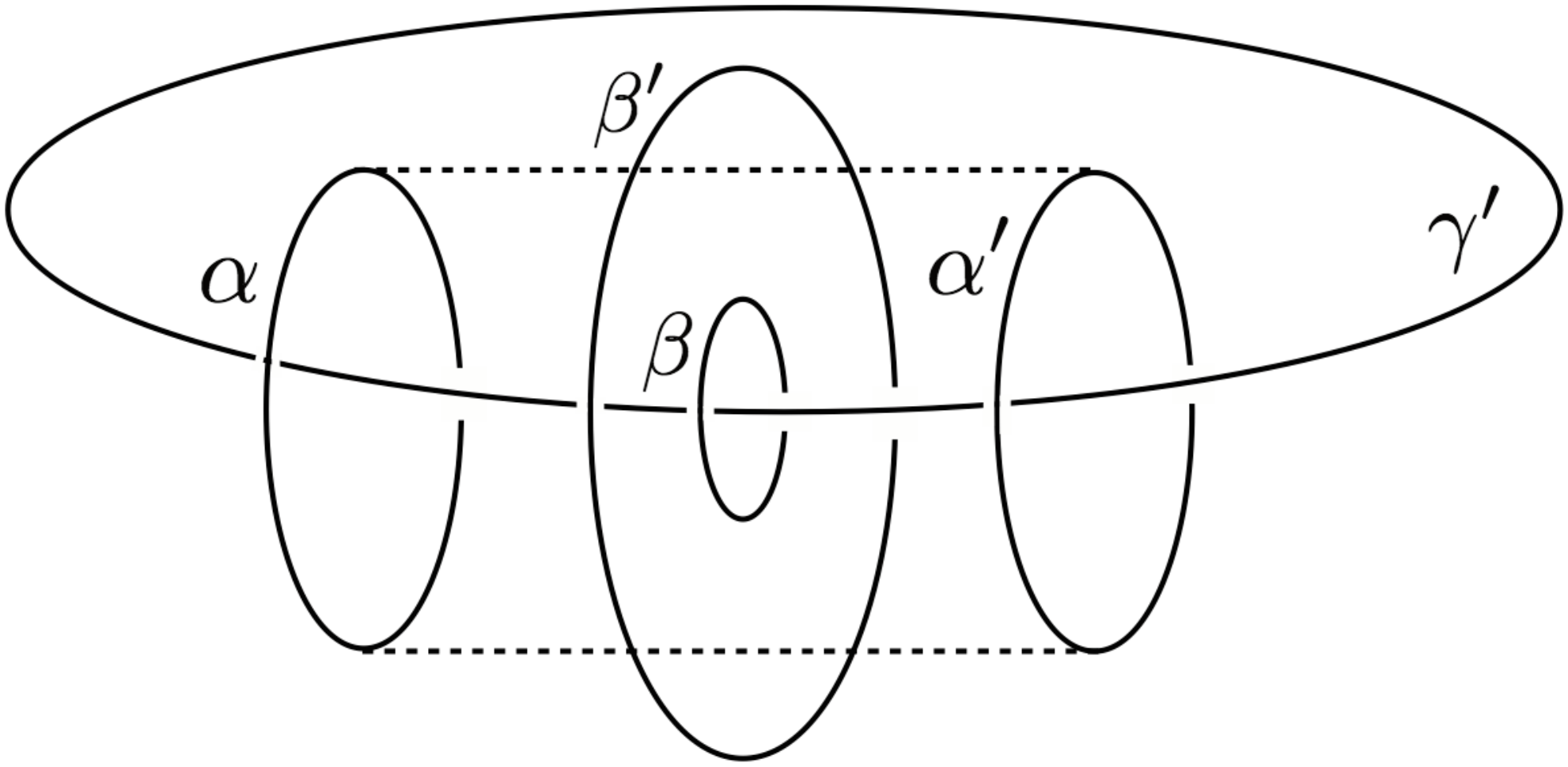}
\caption{Three-loop braiding process as a sequential group commutator $K(K(U_{\alpha}^{\dagger}, U_{\beta}^{\dagger}),U_{\gamma}^{\dagger})$. 
} 
\label{fig_three_braiding}
\end{figure}

The three-loop braiding is depicted in Fig.~\ref{fig_3dim_wall}(c) where one first creates a loop $\gamma$, and then creates/braids $\alpha$ and $\beta$, and then makes the system return to the vacuum. The key distinction from two-loop braiding is that, $\alpha$ and $\beta$ are braided in the presence of another loop $\gamma$. We shall denote the corresponding statistical phase by $e^{i\theta(\alpha,\beta,\gamma)}$. The braiding statistic of three loops $\alpha$,$\beta$,$\gamma$ can be probed by the process depicted in Fig.~\ref{fig_three_braiding}. Let $U_{\gamma}$ be a unitary operator corresponding to creating a pair of loops $\gamma$ and $\gamma'$ from the vacuum and enlarging $\gamma'$. Then, the following sequence of unitary transformations implements the three-loop braiding:
\begin{align}
K(K(U_{\alpha}^{\dagger}, U_{\beta}^{\dagger})^{\dagger},U_{\gamma}^{\dagger})|\psi_{gs}\rangle = e^{i\theta(\alpha,\beta,\gamma)}|\psi_{gs}\rangle,
\end{align}
where 
\begin{align}
K(K(U_{\alpha}^{\dagger}, U_{\beta}^{\dagger})^{\dagger},U_{\gamma}^{\dagger}) = (U_{\alpha}^{\dagger}U_{\beta}^{\dagger}U_{\alpha}U_{\beta})^{\dagger}U_{\gamma}^{\dagger}(U_{\alpha}^{\dagger}U_{\beta}^{\dagger}U_{\alpha}U_{\beta})U_{\gamma}. 
\end{align}
As such, the three-loop braiding statistics corresponds to the vacuum expectation value of the \emph{sequential} group commutator.

Let us now consider the braiding of $m_{AB}$, $s_{BC}$, $m_{CA}$. Recall that $K(U_{\alpha}^{\dagger}, U_{\beta}^{\dagger})$ is a string of $Z$ operators, characterized by edges of color $ACD$. Recall that $U_{\gamma}$ is a membrane of $X$ operators acting on plaquettes of color $BD$. These two operators overlap odd times, and thus the group commutator between $K(U_{\alpha}^{\dagger},U_{\beta}^{\dagger})^{\dagger}$ and $U_{\gamma}^{\dagger}$ leads to a non-trivial statistical phase: 
\begin{align}
e^{i\theta(m_{AB},s_{BC},m_{CA})}=-1.
\end{align}
In general, one finds
\begin{align}
e^{i\theta(s_{K},m_{K'},m_{K''})}=e^{i\theta(m_{K},s_{K'},m_{K''})}=-1,
\end{align}
where $\{ K,K',K''\}=\{ AB,BC,CA\}$, and permutations of $K,K',K''$. For any other combinations of loo-like excitations $\alpha,\beta,\gamma$, one has $e^{i\theta(\alpha,\beta,\gamma)}=1$. 

Finally, we study the braiding statistics of particles and loop-like excitations in a transparent domain wall of the three-dimensional color code in a folded geometry. Let us begin with the braiding with two excitations. The only non-trivial braiding processes are the ones involving $e$ and $m$, such as the braiding of $(e_{A}|e_{A})$ and $(m_{BC}|m_{BC}s_{BC})$. Braiding of electric charges $e_{A}$  and magnetic fluxes $m_{BC}$ on each side of the wall contributes to $-1$ phases in the braiding statistics, which cancel with each other. For the three-loop braiding, the non-trivial processes are the ones involving $a=(m_{BC}|m_{BC}s_{BC})$, $b=(m_{CA}|m_{CA}s_{CA})$, $c=(m_{AB}|m_{AB}s_{AB})$. Due to the symmetry under permutations of color labels, it suffices to study $\theta(a,a,a)$, $\theta(a,a,b)$, $\theta(a,b,a)$, $\theta(a,b,c)$. It is easy to verify $e^{\theta(a,a,a)}=e^{\theta(a,a,b)}=e^{\theta(a,b,a)}=1$. As for $\theta(a,b,c)$, there are two ``$-1$'' contributions to the phase, and thus one obtains $e^{\theta(a,b,c)}=1$. In conclusion, on a transparent $R_{3}$ wall, particles and loops that may condense on the wall possess mutually trivial braiding statistics, in two-excitation and three-loop braiding processes. This result verifies the hypothesis that gapped domain walls and boundaries can be classified by triviality of braiding statistics, including three-loop braiding processes. 

To conclude the paper, we would like to further discuss a relation between braiding processes and group commutators. We have seen that two-loop and three-loop braiding processes are characterized by a group commutator $K(U_{\alpha}^{\dagger},U_{\beta}^{\dagger})$ and a sequential group commutator $K(K(U_{\alpha}^{\dagger}, U_{\beta}^{\dagger})^{\dagger},U_{\gamma}^{\dagger})$. In the $d$-dimensional color code, one can define codimension-$2$ excitations with non-trivial $d$-excitation braiding statistics. This process can be characterized by the following sequential group commutator:
\begin{align}
K( \ldots K( V_{\alpha_{3}},K(V_{\alpha_{2}},V_{\alpha_{1}}))\ldots)
\end{align}
where $V_{\alpha_{j}}^{\dagger}$ corresponds to a unitary operator creating excitations of type $\alpha_{j}$ for $j=1,\ldots,d$. Readers from quantum information science community may find this expression particularly interesting since the sequential group commutator is essential in defining the Clifford hierarchy, the central object in the classification of fault-tolerant logical gates in topological stabilizer codes~\cite{Gottesman98, Bravyi13b, Pastawski15}. Indeed, the non-trivial three-loop braiding statistics in the three-dimensional color code is a direct consequence of the fact that transversal $R_{3}$ operator belongs to the third-level of the Clifford hierarchy. These observations may hint a deeper connection between classification of fault-tolerant logical gates, braiding statistics in higher-dimensional TQFTs and quantum error-correcting codes. It will be also interesting to study logical gates and boundaries in topological quantum codes which are beyond TQFTs~\cite{Haah11, Beni13}.

\section*{Ackowledgment}

I would like to thank Xie Chen, Isaac Kim, Aleksander Kubica, Iman Marvian, Fernando Pastawski, John Preskill and Sujeet Shukla for discussions and/or comments. I would like to thank Aleksander Kubica for careful reading of the manuscript. Part of the work was completed during the visit to the Perimeter Institute for theoretical physics. I am supported by the David and Ellen Lee Postdoctoral fellowship. I acknowledge funding provided by the Institute for Quantum Information and Matter, an NSF Physics Frontiers Center with support of the Gordon and Betty Moore Foundation (Grants No. PHY-0803371 and PHY-1125565). 


\end{document}